\documentclass[preprint]{aastex}
\usepackage{psfig}

\received{}
\revised{}
\accepted{}
\ccc{}
\cpright{}{}

\slugcomment{Accepted for publication in ApJ}
\shorttitle{}
\shortauthors{}

\newcommand{\um}{$\mu$m}
\newcommand{\spitzer}{\textit{Spitzer}}

\begin{document}



\title{Spectacular {\it Spitzer} Images of the Trifid Nebula:
Protostars in a Young, Massive-star-forming Region}

\author{Jeonghee Rho and William T. Reach}
\affil{{\it Spitzer} Science Center,
California Institute of Technology, Pasadena, CA 91125; rho@ipac.caltech.edu, reach@ipac.caltech.edu}

\author{ Bertrand Lefloch} 
\affil{Laboratoire d'Astrophysique, Observatoire de Grenoble, BP 53,
F-38041, Grenoble CEDEX9, France; lefloch@obs.ujf-grenoble.fr}

\author{Giovanni G. Fazio} \affil{Harvard-Smithsonian Center for Astrophysics, 
MS 42, 60 Garden Street, Cambridge, MA 02138; gfazio@cfa.harvard.edu }

\begin{abstract}

\spitzer\ IRAC and MIPS images of the Trifid Nebula (M20)  reveal its
spectacular appearance in infrared light, highlighting the nebula's
special evolutionary stage. The images feature recently-formed  
massive protostars and numerous young stellar objects, and a single O star that
illuminates the  surrounding molecular cloud from which it formed, and
unveil large-scale, filamentary dark clouds. The  hot dust grains show
contrasting infrared colors in  shells,  arcs, bow-shocks and dark
cores.  Multiple protostars are detected in the infrared, within the
cold dust cores of TC3 and TC4, which were previously defined
as Class 0. 
The cold dust continuum cores of TC1 and TC2
contain only one protostar each; the newly discovered  infrared source
in TC2 is   the driving source of the HH399 jet.   The \spitzer\
color-color diagram allowed us to identify $\sim$160 young stellar
objects (YSOs) and classify them into different evolutionary stages.
The diagram also revealed a unique group of YSOs which are bright at
24$\mu$m but have the spectral energy distribution peaking at
5-8$\mu$m.   Despite expectation that Class 0 sources would be
``starless" cores,   the \spitzer\ images, with unprecedented
sensitivity,  uncover mid-infrared emission from these Class 0
protostars.  The mid-infrared detections of Class 0 protostars show
that the emission  escapes the dense, cold envelope of young
protostars. The mid-infrared emission of the  protostars can be fit
by two temperatures of 150 and 400 K; the hot core region 
is probably optically thin in the mid-infrared regime, and  the size of hot
core is much smaller than that of the cold envelope.
The presence of multiple
protostars within the cold cores of Class 0 objects implies that
clustering occurs at this early stage of star formation. 
The TC3 cluster shows that the most
massive star is located at the center of the cluster and
at the bottom of the gravitational-potential well. 

\end{abstract}


\section{Introduction}
The Trifid Nebula (M20), is one of the best-known astrophysical objects:
a classical nebula of ionized gas from an O7 star (HD 164492). The nebula
 glows  in
red light, trisected by obscuring  dust lanes, with a reflection nebula
in the north.  The Trifid Nebula is a very young  \ion{H}{2} region with an
age of $\sim 3\times 10^{5}$ years. The Infrared Space Observatory
(ISO) and the Hubble Space Telescope (HST) (Cernicharo et al.\ 1998;
Lefloch \& Cernicharo 2000; Hester et al.\ 1999) show the Trifid to be
a dynamic, ``pre-Orion" star forming region containing young stellar objects (YSOs)
undergoing episodes of violent mass ejection, and protostars (like
HH399) losing mass and energy to the nebula in jets. 
Four massive (17--60 M$_\odot$) protostellar cores were  discovered, with
millimeter-wave observations, in the Trifid (Lefloch \& Cernicharo
2000). A number of YSO candidates were
identified  using near-infrared  (Rho et al. 2001) and X-ray
observations (Rho et al. 2004). 
 We adopted a distance of 1.68 kpc for 
the Trifid Nebula, which was measured by Lynds et al. (1986).
There have been two primary tools for studying massive star formation:
molecular line emission which traces dense material and outflows  (e.g.
Shirley et al. 2003), and dust continuum emission which traces cold,
dense material (e.g. William et al. 2004).  The youngest objects,
Class 0 protostars, are those
that are detected as condensations in dust continuum
maps characterized by very low values of the ratio L$_{bol}$/L$_{submm}$, 
and show collimated CO outflows or internal-heating sources
(Andr\'e 1994). The Class 0 protostars were believed to
be ``starless" cores,  with neither near-infrared nor
mid-infrared ($<$20$\mu$m) emission. In this paper, we 
present spectacular \spitzer\ images of the Trifid Nebula and report $\sim$160  
newly identified young stellar objects (YSOs)  using their infrared
colors. We  found  multiple protostars within cold cores and
many evolved YSOs are located along the ionization fronts. We illustrate
that  \spitzer\ infrared images  provide new, excellent tools for
studying massive-star formation in ways  that were not previously
available.

\section{Observations}
We performed our infrared observations of the Trifid Nebula with the {\it Spitzer} Space
Telescope, the fourth NASA Great Observatory.
The Infrared Array Camera (IRAC; Fazio et al. 2004)  and
Multiband Imaging Photometer  for {\it Spitzer}  (MIPS; Rieke et al.
2004) observations took place on March 31, 2004  (AOR = 6049024) and
April 11, 2004  (AOR = 6048768), respectively.  The integration time 
is 8 seconds for IRAC and  48 seconds for MIPS per sky position. The data
are processed with the \spitzer\ Science Center (SSC)
pipeline\footnote{http://ssc.spitzer.caltech.edu/irac[mips]/documents}.
For individual frame data known as ``Basic Calibrated Data" (BCD),
outlier rejection was applied  to remove cosmic rays, and dark and flat fields 
were also applied. The full-width-half-maximum (FWHM) of the 
point-spread-function (PSF)  is approximately 1.7$''$ in the IRAC images and  6$''$
in the  MIPS 24\um\ image. The uncertainty of the calibration  is less
than 5\% for all IRAC wavelengths and  less than 10\% for the MIPS
24\um. The photometry for the sources was performed using aperture photometry, IRAF APPHOT
(we used aperture sizes of  4$''$, 4$''$, 5$''$,  6$''$ and 14.7$''$
for IRAC bands 1, 2, 3, 4 and MIPS 24$\mu$m, respectively)   and the
photometric errors are  less than 0.1 mag; when the sources are located
at bright diffuse emission,  the photometric uncertainty are higher. 
The zero-magnitude flux
densities 
are 280.9, 179.7, 115.0, 64.13 Jy for IRAC bands 1, 2, 3, and 4, respectively 
(Reach et al. 2005), and 7.15 Jy for    
MIPS 24$\mu$m.
The number of
extracted sources (S$/$N$>$ 5) is 21,400 at 3.6\um, 2,665 at 8\um,
and 541 at 24$\mu$m.

 We also performed follow-up near-infrared
observations using the Palomar 200-inch Wide-infrared Camera (WIRC).
The data were taken  on Aug 23-26, 2004 and the exposure time was 60 sec per
sky position. 
The sensitivities in near-infrared observations  are
$\sim$20.0 mag for J and H bands and $\sim$17 mag for K$_s$ band. 
The photometry at all wavelengths was merged into a 
catalog for sources detected at 8\um.

\section{{\it Spitzer} infrared images and the color-color diagram }

The \spitzer\ IRAC and MIPS images of  the Trifid Nebula (see Figure 1)
reveal its
spectacular appearance in infrared light, highlighting the nebula's
special evolutionary stage. The images feature recently-formed  
massive protostars and numerous young stellar objects, and a single O star that
illuminates the  surrounding molecular cloud from which it formed, and
unveil large-scale, filamentary dark clouds. The  hot dust grains show
contrasting infrared colors in  shells,  arcs, bow-shocks and dark
cores. 
  The 8$\mu$m emission
represents polycyclic aromatic hydrocarbon (PAH) structures  
coinciding with the optical dust lanes and shows stripped clouds 
evaporating through strong UV radiation of the exciting O star, indicating 
that the PAH is only partially destroyed in \ion{H}{2}
regions.  Numerous young stellar objects are detected as apparent color excesses
in red or yellow.

By using multi-wavelength photometries,  we generated  the color-color
diagram  shown in Figure~\ref{tricolcol} and the  spectral  energy
distributions (SEDs) of the infrared excess stars. In the color-color
diagram,  main sequence stars fall near (0,0) magnitudes.  The
protostars (Class I/0) are the reddest sources, toward the
upper-right of Figure~\ref{tricolcol},  and young stellar objects with disks
(Class II) are intermediate.  This color-color classification is based
on infrared colors of YSOs (Reach et al. 2004). We also confirmed
the color-color classification with the full SEDs of each category of
YSOs. In the SEDs, Class I/0 SEDs rise  from near-infrared to 24\um, while Class II
SEDs decrease in the near-infrared but  show  infrared excesses in
comparison with the photosphere, as shown in Figure \ref{classsed}.  A
large population of young stellar objects is present as listed in Table 1: 
37 in Class I/0, and  111
in Class II.   The extinction vector is shown in the color-color
diagram (Figure ~\ref{tricolcol}).  The extinction towards Trifid
is only A$_v$=1.3 \citep{lyn85}, so it can only change  the colors of
[8]-[24]=0.04 mag and [3.6]-[5.8]=0.02 mag, which is too small  to affect
the classification. 
However, considering the local extinction towards YSOs,
classification of the objects which fall at the boundaries 
between the  Class 0/I and Class II objects
in Fig. \ref{tricolcol} may
be somewhat uncertain.


A unique group of YSOs  with [8]-[24]=1.5$\sim$2.5 and
[3.6]-[5.8]$>$1.5 appears in the \spitzer\ color-color diagram; their
SEDs peak  between 3 and 8\um\ and typically they  are bright 24\um\
sources. We name them ``Hot excess" young stellar objects, because of
the extra, hot ($\sim$500K temperature) component.  We note that one of
these sources is a B-type emission line star, LkH$\alpha$ 124,
indicating ``Hot excess" stars are  YSOs. The ``Hot excess" stars could
mostly  be Herbig Ae/Be stars; such stars are believed to be the
intermediate mass analogue of T Tauri stars. 
In order for Class II objects to be ``Hot excess"  YSOs,  a high
extinction (A$_v$$>$25 mag) is required.
Another possibility is the ``Hot excess" YSOs are  Class I/0 protostars
with an extra  hot component from more active accretion than typical,
in addition to the cold envelope. We have noticed  that the number of
``Hot excess" YSOs in the Trifid Nebula is higher than those in other
star-forming regions (an example is  Figure 2 of Muzerolle et al.
2004); this may be related to the fact that the Trifid Nebula is in a
dense star-forming environment with a high UV-radiation-field.
Far-infrared observations would be needed to unambiguously identify the
evolutionary stage of the ``Hot excess"  YSOs.  

The supergiant star HD 164514, A7Iab, is responsible for the blue optical 
reflection nebula on the northern side of the Trifid Nebula, and is not an young stellar
object.
This star is located at the top left in the color-color diagram ([8]-[24]$<$1.5
and [3.6]-[5.8]$>$1.5) in the region marked as ``Giants" in Figure \ref{tricolcol}; 
note that it is located far to the
left of ``Hot excess" stars.
Other sources at the same
location in the color-color diagram may be similar objects because of
similar infrared colors.  
We also examined the sources with very high color excesses in
[8]-[24] ($>$5) but no color excess in  [3.6]-[5.8] (=0 to 1.5), and
found that each of the objects has patches of extended emission such as
reflection/emission nebulae.

We compared the distribution of YSOs in the \spitzer\ images  with  the
cold dust continuum map \citep{cer98}. Most of the protostars (Class I/0)
are distributed along  the dark filamentary cloud on the western side of
M20 (see Figs. ~\ref{trispicolor} and ~\ref{tc3tc4spitzer})  and along
the optical dust lanes that trisect the \ion{H}{2} region.   Most of the
Class II stars, on the other hand, are distributed along the ionization
fronts of the Trifid Nebula, following the circular shape of the \ion{H}{2} region.   While
protostars coincide with the  dust continuum cores, TC1 to TC4, they are
not always  precisely  correlated with the dust emission peaks.

Just north of the Trifid Nebula is a region containing a surprising
number of protostars  (R.A.\ $18^{\rm h}  02^{\rm m} 42.28^{\rm s}$ and
Dec.\  $-22^\circ$48$^{\prime} 44.8^{\prime \prime}$, J2000). Both the
diffuse mid-infrared emission and the distribution of young stellar objects  in
the 8$\mu$m and 24$\mu$m maps suggest that this is a small \ion{H}{2} region
with active star formation. A heating source is indicated by the
presence of bright 24\um\ emission from hot dust grains, and PAH-rich
dust lanes surround the nebula as in the Trifid proper. We  analyzed a
20cm VLA map \citep{yus00}, and 
adopting an electron  temperature of $10^4$ K and a recombination
coefficient  of  2.7$\times 10^{-13}$ cm$^3$ s$^{-1}$,
we found  an  integrated  flux of 0.8 Jy
and   a radiative recombination rate of 10$^{47.23}$ s$^{-1}$.
The  ionization
rate is consistent with  a young B0/O9.5 star or an evolved B0.5 (giant
or supergiant). Hence, we suggest a massive star is hidden in the area,
which implies that the region north of the Trifid Nebula proper is a separate
\ion{H}{2} region, which we named  ``Trifid Junior."

\section{Mid-infrared Emission from Protostars and  
Multiple-protostars within the dust continuum cores}

We identified 5 protostars within the  dust continuum
cores of TC3 and TC4; they are listed in Table 2 and shown
in Figures \ref{hstspitzer} and \ref{tc3tc4spitzer}. 
The {\it Spitzer}  color-color diagram shows that most of the infrared sources are Class
I/0 protostars and two sources are ``Hot excess" stars. 
Figure
\ref{tc3tc4spitzer} shows three color {\it Spitzer} images superposed
on dust continuum contours. The TC3 and TC4 condensations have been 
classified as Class 0 protostellar cores.  Previously, ISO
observations at  70$''$ resolution implied the presence of a
cold component at temperatures of $\sim$20 K for both
TC3 and TC4 \citep{lef00}.

We constructed  the spectral energy distribution of each infrared
Class I/0 protostar within the TC3 and TC4 cores,  from 1.25--24$\mu$m combining
the Palomar, IRAC, and MIPS 24$\mu$m data. The luminosities are derived
from the SEDs  between 1.25--24$\mu$m and are listed in Table 2. All
protostars show rising SEDs from near- through mid-infrared.
Figure~\ref{sedtc43} shows the SEDs of a few protostars within the TC4 and
TC3 cores. 
The depression in the 8$\mu$m flux relative to the apparent continuum
at 5.8 and 24 $\mu$m, is noticeable particularly for TC3A, TC3E, TC4A
and TC4D. This is likely due to  presence of deep silicate absorption. 
By using multi-color-temperature fits to the SEDs, we  estimated the
luminosities of TC4 and TC3. Here we added millimeter and 160$\mu$m
fluxes; the dust continuum sources TC4 and TC3 have 
millimeter fluxes of 0.6 and 0.92 Jy integrated over the half-power contours, respectively \citep{lef00}. 
The mid-infrared fluxes for TC3 and TC4
include the sum of TC3A-TC3C and of TC4A-TC4B, respectively, because
they are within  the 50\% millimeter contours of TC3 and
TC4 cores. We also
generated the SED of TC4A,  the brightest infrared source in the TC4 core,
assuming a
significant amount of the 1.3mm flux is from Component A.
To fit this SED, we use multiple components: a cold envelope with temperature
and optical depth determined by the 160--1250$\mu$m data, and warm component
mostly required by the 24$\mu$m point, and two hot components to match the
near to mid-infrared points. The warm and hot components are extinguished
by the envelope, but are probably optically thin themselves (see section 5 for details). 
The SEDs of TC3, TC4 and TC4A,
have similar best-fit temperatures of T$_{cold}$ = 22 K, T$_{warm}$ = 150
K, T$_{hot1}$ = 400 K, and  T$_{hot2}$ = 1300 K. An example SED of TC4A is
shown in  Figure \ref{sedtc4a}.  The SEDs of the TC4 and TC3 cores when
adding the mid- and near-infrared fluxes, yielded total luminosities
of  $\sim$1200 L$_{\odot}$ and 1700 L$_{\odot}$, respectively.  This
implies that the protostars are moderately massive.   The estimated luminosity of
TC4 falls in the range of 500-2000 L$_{\odot}$, which was previously
estimated  by \cite{lef00}.  The near- to mid-infrared  luminosities
of the infrared protostars range from 1 to 32
 L$_\odot$ as listed in Table 2. The warm and hot temperature
components contribute less than a few percent of the total luminosity.
However, the mid-infrared detection from the earliest protostars is
new, which is important for understanding of their SEDs.

The MIPS 24$\mu$m data points are fit by a temperature of  $\sim$150
K; we suggest that this emission originates from the warmer, inner part
of the cold envelope. Most of the IRAC data points are fit by a
temperature of  $\sim$400 K. The SEDs indicate that the protostars  are
hot cores within cold, dense envelopes and hot cores require internal
heating.   For TC4A, TC4C, and TC3D, the near-infrared emission  
requires a color temperature of $\sim$1300 K;  the other sources in
Table 2 have one or two band detections in the near-infrared and  their
SEDs do not require the presence of the 1300 K temperature component. 
Our detection of the extremely hot temperature of 1300 K, close to the dust sublimation
temperature (the exact sublimation temperature  depends on  particular
dust properties), is the first one from the earliest-type 
protostars as far as we are aware.  TC4C and TC3D were classified as $``$Hot excess" stars so
they may be  in somewhat advanced stages like  Class I or II. However,
it is surprising to find  near-infrared emission from TC4A, the
brightest infrared protostar in the TC4 core,  which is believed to be
a Class 0 object \citep{lef02}.   Some possible explanations are as
follows. First, TC4A may have a star inside the envelope; in other
words, nucleosynthesis may have already begun in the core. This implies
that there is no clear evolutionary distinction between Class 0 and
Class I.  Another possibility is  that the near-infrared emission is
light scattered directly from the central source. Recent theoretical
simulations by \citet{ind06} showed that near-infrared emission from
protostars is a few times brighter  when the envelope has a clumpy 
medium, due to scattered-light from the central source.  When we
examined our near-infrared images, we noticed that  some of the
protostars may have extended near-infrared emission around the
point-like sources. The narrow-filter images  did not show any excess
emission  such as in  H$_2$ and [Fe II]; this result supports the 
hypothesis that the near-infrared emission may be scattered-light
rather than a powering source of a jet or outflow.  We also cannot rule
out the possibility that near-infrared counterparts are from low mass
companions. However, we favor the hypothesis that the near-infrared emission
is scattered-light directly from the central source over a possibility
of a star inside the cold envelope or from companions. 

The mid- and near-infrared luminosities (L$_{mid-IR}$) were estimated
using the fits to the SEDs  from 1.2 to 24$\mu$m fluxes,  as shown in
Table 2.  TC3A, the brightest 24$\mu$m protostar in the TC3 core, is
located in the millimeter peak and at the center of the core. However,
the brightest 24$\mu$m source in the TC4 core, TC4A, is slightly off
center. The second dust continuum peak in the TC4 core 
(TC4b in Lefloch \& Cernicharo 2000) dust continuum peak 
contains the \spitzer\ infrared sources of TC4D and TC4E, and
has an envelope mass of 11 M$_{\odot}$\citep{lef00}. Five
protostars in the TC4 core have mid-infrared luminosities
ranging between 6-32 L$_{\odot}$,  and the protostars in the TC3 core
have between 1-16 L$_{\odot}$. This suggests that
protostars within the TC3 and TC4 cores are composed of a mixture of
different-mass YSOs.

In contrast to TC3 and TC4 systems, the dust continuum cores of TC1 and
TC2, contain only one protostar for each, as listed in Table 2 and
shown in Figure \ref{hstspitzer}.  The infrared protostars were
identified as  Class I/0 by using the color-color diagram.  A   heated
dust shell appears around the protostar in the TC1 core, as shown in
Figure ~\ref{hstspitzer}, and the infrared shell  surrounds the dark
cold envelope.  The images of the TC2 dust core unveil a point source
lying along the axis of the HH~399 jet (see also Fig.~\ref{hstspitzer})
and the point source has typical protostellar colors. The SEDs of the
protostars of the TC1 and TC2 infer that both bolometric luminosities
are $\sim$600 L$_{\odot}$ and the mid-infrared luminosities are 19 and 24 
L$_{\odot}$ for
TC1 and TC2, respectively.

\section{Discussion}

The detection of multiple mid-infrared protostars from the cold dust cores is a new result. 
How does the mid-infrared emission leak through  thick envelopes of
material infalling onto the stars? Isn't the mid-infrared emission
absorbed by the cold and  thick envelopes?  We can address where the
emission arises, how the dust is heated, and how the emission escapes
cold cores.  The cold envelope sizes of the TC3, TC4, and TC4b cores are  0.2, 0.2,
and 0.16 pc, respectively , which were  determined  by \citet{lef00},
using the 50\% contour in the millimeter flux map. The estimated
optical depth is:
\begin{equation}
 \tau_{1250} = {F_{1250} \over {\Omega \, 
B_{\nu}(T)}} = 1.08\times 10^{-4}  F_{1250}
{d_{1.68kpc}^2 \over  R^{2} }
\label{eq1}
\end{equation}

\noindent where F$_{1250}$  is the mm-wave flux,
$\Omega$ is a solid angle of  the source size,   B$_{\nu}$(T) is a
black body Planck function at a temperature of T,  and R is a source
radius in unit of pc.  The optical depths of the TC3, TC4, and TC4b cores
are between  0.7-2.5 $\times 10^{-3}$ (for a temperature of 22 K). 
The corresponding optical depths at 8$\mu$m were determined using
$\tau_{8}$ =
$\tau_{1250} \times (1250/8)^{\alpha_1}$ $(8/100)^{\alpha_2}$, where
$\alpha_1=2$ is the extinction slope between 100-3000$\mu$m and
$\alpha_2=1$  is the extinction slope between 10-100$\mu$m (Li
\& Draine 2001); this yields 
$\tau_{8 \mu m}$ $\sim$ 1--7.
The corresponding visual extinction was determined using
A$_{8{\mu}m}$/A$_K$ = 0.5 (Draine 2003); this yields 
a total extinction of A$_v$  $\sim$ 20--150.

Despite the high extinction of the envelope and the high optical depth
of the protostars (estimated using a temperature of 22K) at 8$\mu$m, 
apparently we observed mid-IR emission from the early protostars. We
note that the optical depths of 150 K and 400 K components are much
smaller, more than one magnitude smaller than that of the cold (22 K) envelope.
 From the SED fits (see  Fig. \ref{sedtc4a}) we estimated that  the
optical depth of the 8$\mu$m emitting region is $< 10^{-3}$ for the warm and hot
temperature components, which suggests that the accretion region is probably optically thin
if it is larger than $\sim 10$ AU.
In addition,
the \spitzer\ images revealed that the infrared sources of the protostars
were point-sources, inferring that the regions emitting 400 K and 150 K
would be smaller than  the PSF radii of the IRAC and MIPS 24$\mu$m images,
which are 0.007 pc (= 1500 AU) and 0.025 pc (= 5$\times 10^3$ AU), respectively.
This is consistent with
the models of the massive protostars \citep{oso99};
the cold component (responsible for far-infrared and 1300$\mu$m
emission) would be emitted from a region $>$ $10^4$ AU  from the  central
source, while the 400 K component (responsible for mid-infrared
emission) would be  from $<$ 1500 AU region. 
The mid-infrared emission
can escape the dense envelope; it suffers a modest mid-infrared
extinction  but traces the emission directly from the accretion
region.  Previous ISO observations showed similarly  mid-infrared
counterparts of a few low mass Class 0 objects in other star forming
regions \citep{cer00}. 
New \spitzer\ observations of Class 0 objects also detected
mid-infrared protostars in L1014 (Young et al. 2004) and  Cepheus E
\citep{nor04}. The detections infer that  the region of
hot core in early protostars is  optical thin at  the mid-infrared regime,
and the size of hot core is smaller than the cold envelope.

The mid-infrared emission is powered by accretion of envelope material
onto the protostar, and  most of the bolometric luminosity at the early stage
of the protostar is
from accretion. The accretion rate is given by \cite{mck02,mck03}:

$$\dot{m_{*}}= 4.75\times10^{-4}\,\, \epsilon_{core}^{1/4} (f_{gas}
\,\phi_P \,\alpha_{vir})^{3/8} \,\,  \biggl({m_{*f} \over {30
M_\odot}}\biggr)^{3/4} \Sigma_{cl}^{3/4} \biggl({m_{*} \over m_{*f}
}\biggr)^{1/2} \,\, M_{\odot} \, yr^{-1} $$ 
\begin{equation}
 \sim 4.6\times 10^{-4}  \biggl({m_{*f} \over {30
M_\odot}}\biggr)^{3/4} \Sigma_{cl}^{3/4}  \biggl({m_{*} \over m_{*f}
}\biggr)^{1/2} \,\, M_{\odot} \, yr^{-1}  
\label{eq2}
\end{equation}

where $\epsilon_{core}$ is
the fraction of the total core mass (M$_{core}$),   $\phi_P$ is the
ratio of the core's surface pressure to the mean pressure in the
clumps, $f_{gas}$ is the fraction of the cloud's mass that is in gas,
as opposed to stars, $\Sigma_{cl}$ is a mean mass column density,
$m_{*}$ (=$\epsilon_{core}$M$_{core}$, where M$_{core}$ is the total
core mass) is  the instantaneous mass (the  current mass), and $m_{*f}$
is the final mass of the star. We used the values of $\Sigma_{cl}$ =1,
$\epsilon_{core}$=0.5  and $\phi_P$=0.663 (see \cite{mck02}
for details). The bolometric luminosity is the sum
of the internal and accretion luminosity (L$_{bol}$ = L$_{int}$ +
L$_{acc}$). The accretion luminosity is given by \cite{mck03}:
\begin{equation}
 L_{acc} \sim 3.0\times10^4 \,\, \biggl({f_{acc} \over 0.5}\biggr)
\biggl({m_{*f} \over {30 M_\odot}}\biggr)^{1.2}  \Sigma_{cl}^{3/4}
\biggl({m_{*} \over m_{*f} }\biggr)^{0.95} \,\, \Sigma_{cl}^{3/4}  \,\,
L_{\odot}  
\end{equation}
where
${f_{acc}}$ ($<<$1) is a factor that accounts for the energy advected
into the star or used to drive protostellar outflows.

The internal luminosity, L$_{int}$, is equal to the luminosity
transported by radiation and is determined mostly by the stellar mass
(L$_{int}$ $\propto$ M$_*^{5.5}$ R$_*^{-0.5}$ $\propto$ M$_{*}$). A
simple formula cannot be used for a wide range of masses of stars,
because as the mass increases, the relative contribution of the Kramers
and electron scattering opacities changes \citep{nak00}. Therefore,
in order to estimate the mass of star from the bolometric luminosity,
we directly used  Figure 2 of \citet{mck02}. 
The instantaneous masses of TC3A and TC4A are 3-8
M$_{\odot}$ and 2-5 M$_{\odot}$ for the bolometric luminosities of
$\sim$ 1700 and 1200 L$_{\odot}$, respectively, assuming a
significant amount of the 1.3mm flux is from Component A.  The
accreting properties of the TC3 core are similar to those of G34.24+0.1
\citep{mck03}.  We estimated accretion rates using the equation 2 
for a grid of final masses of 7.5, 30 and 120 M$_\odot$.  The accretion
rates for TC3 and TC4 are 1-5$\times$10$^{-4}$ M$_{\odot}$ yr$^{-1}$
and 0.9-4$\times$10$^{-4}$ M$_{\odot}$ yr$^{-1}$, respectively. 
We also estimated the current masses and the accretion rates of TC1 and TC2 
 using L$_{bol}$ of 600 L$_{\odot}$;
the current masses are 1-3 M$_{\odot}$ and the  accretion rates
are 0.8-1.7$\times$10$^{-4}$ M$_{\odot}$ yr$^{-1}$. 
The estimated accretion luminosities infer that high portions of the bolometric
luminosities are from accretion luminosities in the early stage of 
massive-star formation. The timescale of the infalling stage is determined
mainly by  the conditions in the stars' natal cloud and weakly depends
on the mass of stars.

In the study of massive-star formation, a fundamental open question is
how clusters are formed.  Do massive dense cores have internal
substructures? Do clumps evolve independently to produce stars, or do they share
a common evolutionary process?  Are apparent-single stars born
single, or are they born in groups and subsequently ejected? While
low-mass stars are believed to be produced mainly through accretion (Lada
1991), there are  two main scenarios to explain  high-mass star
formation.  One scenario is through accretion,  like that of low-mass stars but
with higher accretion rates; the other scenario is through formation
of  high-mass stars through coalescence of lower-mass protostars.
We found  5 protostars for each of TC3 and TC4 cores, and 
the brightest and the most massive star in each cluster is located in
the dust continuum peak (see Table 2), implying these systems
are possible protoclusters. The brightest protostar appears near the
center of the dust continuum peak. The brightest 24$\mu$m source in
the TC3, TC3A, is located in the millimeter peak and at the center of
the core.  Within the cluster, as the separation of TC3B-TC3C
from the largest star
of each Component A increases, the mid-infrared luminosity  
decreases in the case of the  TC3 core, as shown in Table 2.
This suggests that the brightest and most massive star among 4
protostars  is  located at the bottom of the gravitational
potential well and is located where it formed,  although the
mid-infrared luminosities may not be directly correlated with the total
bolometric luminosities.  Accretion from gas or nearby  fragmented clumps
is found through the accretion of residual gas onto relatively 
low-mass cores. This picture may be consistent with  the mass segregation,
in young stellar clusters,  and competitive accretion proposed by
Bonnell \& Davies (1998) and Bonnell et al. (2004). 
However, the
brightest 24$\mu$m source in the TC4 core, TC4A, is slightly off
from the center of the millimeter peak, although TC4A is
still the largest star in the cluster and the closest star to the dust continuum peak.
The mid-infrared
luminosity of TC4B is also smaller than that of the TC4A.
It is unknown how these mid-infrared luminosities correlate with the total
bolometric luminosities for individual sources. 
Comparable
resolution far-infrared or millimeter observations are required in order to
answer this question. Unfortunately, 
even through the Trifid Nebula is one of the nearest massive-star-forming regions with
a distance of only 1.68 kpc, protostars in the Trifid as well as other comparable
massive-star-forming regions
are not bright enough for such observatories as
the submillimeter Array (SMA). Future far-infrared observations with comparable
spatial resolution are
needed to answer this question.

\acknowledgements
We thank those who have been involved in the successful \spitzer\ mission and
SSC colleagues who participated in the initial Early Release proposal of M 20.
We thank Tom Jarrett for help in analyzing the Palomar/WIRC data.
J. Rho thanks Susana Lizano for helpful discussion of early protostars.
We thank Robert Hurt for producing Figures \ref{trispicolor}
and \ref{hstspitzer} for our press release.
This work is based on observations made with the \spitzer\ {\it Space Telescope},
which is operated by the Jet Propulsion Laboratory, California Institute of Technology,
under NASA contract 1407. Support for this work was provided by
NASA through an award issued by JPL/Caltech.

\begin{deluxetable}{llllccccccccc}
\tabletypesize{\scriptsize}
\rotate
\tablenum{1}
\tablewidth{0pt}
\tablecaption{ Protostars and Young Stellar Objects in the Trifid Nebula
identified by the {\it Spitzer} color-color diagram}.
\label{catalog}
\tablehead{
\colhead{Class No.}    &
\colhead{   SSTM20  }    &
\colhead{   MIPS 24 }    &
\colhead{   IRAC 8   }    &
\colhead{   IRAC 5.8  }    &
\colhead{  IRAC 4.5 }    &
\colhead{IRAC 3.6  }    &
\colhead{K\tablenotemark{a}}    &
\colhead{H\tablenotemark{a} }    &
\colhead{J\tablenotemark{a}}   \\
\colhead{    }       & 
\colhead{   (J2000)  }              &
\colhead{    (mag) } &
\colhead{    (mag)} & 
\colhead{    (mag)} & 
\colhead{    (mag)} & 
\colhead{    (mag)} &     
\colhead{(mag)} & 
\colhead{(mag)}& 
\colhead{(mag)} 
 }     
\startdata
ClassI0  1&18:01:53.69-23:09:54.4& 1.83$\pm$ 0.01& 5.49$\pm$ 0.01& 7.08$\pm$ 0.01& 9.74$\pm$ 0.03&10.46$\pm$ 0.03&14.71$\pm$ 0.03&17.10$\pm$ 0.12&99.99$\pm$ 9.99&\\
ClassI0  2&18:01:55.11-22:56:38.4& 5.11$\pm$ 0.05& 8.36$\pm$ 0.03& 9.37$\pm$ 0.04&10.21$\pm$ 0.04&10.92$\pm$ 0.04&14.26$\pm$ 0.02&16.66$\pm$ 0.12&99.99$\pm$ 9.99&\\
ClassI0  3&18:01:57.72-22:55:12.0& 4.07$\pm$ 0.03& 8.08$\pm$ 0.03& 9.01$\pm$ 0.03&10.48$\pm$ 0.04&11.62$\pm$ 0.12&17.11$\pm$ 0.17&99.99$\pm$ 9.99&99.99$\pm$ 9.99&\\
ClassI0  4&18:02:01.77-23:05:53.9& 3.03$\pm$ 0.02& 7.13$\pm$ 0.02& 7.71$\pm$ 0.01&99.99$\pm$ 9.99& 9.54$\pm$ 0.02&12.91$\pm$ 0.02&15.73$\pm$ 0.05&19.09$\pm$ 0.18&\\
ClassI0  5&18:02:04.68-23:00:15.5& 2.77$\pm$ 0.09& 7.34$\pm$ 0.04& 8.60$\pm$ 0.03&10.04$\pm$ 0.03&10.21$\pm$ 0.03&12.03$\pm$ 0.02&99.99$\pm$ 9.99&99.99$\pm$ 9.99&\\
ClassI0  6&18:02:05.45-23:05:05.3& 4.36$\pm$ 0.10& 7.92$\pm$ 0.12& 8.71$\pm$ 0.05& 9.57$\pm$ 0.02&11.34$\pm$ 0.05&16.49$\pm$ 0.08&99.99$\pm$ 9.99&99.99$\pm$ 9.99&\\
ClassI0  7&18:02:05.57-23:05:29.0& 2.93$\pm$ 0.03& 7.88$\pm$ 0.08& 8.45$\pm$ 0.03& 9.66$\pm$ 0.02&11.18$\pm$ 0.04&99.99$\pm$ 9.99&18.04$\pm$ 0.40&99.99$\pm$ 9.99&\\
ClassI0  8&18:02:07.20-23:05:36.2& 3.94$\pm$ 0.08& 7.18$\pm$ 0.07& 7.54$\pm$ 0.03& 8.30$\pm$ 0.01& 9.44$\pm$ 0.02&14.43$\pm$ 0.02&99.99$\pm$ 9.99&99.99$\pm$ 9.99&\\
ClassI0  9&18:02:08.35-22:44:55.0& 2.46$\pm$ 0.03& 6.80$\pm$ 0.01& 8.34$\pm$ 0.02&10.29$\pm$ 0.06&10.73$\pm$ 0.04&99.99$\pm$ 9.99&99.99$\pm$ 9.99&99.99$\pm$ 9.99&\\
ClassI0 10&18:02:09.46-22:47:14.6& 3.98$\pm$ 0.06& 7.17$\pm$ 0.03& 7.80$\pm$ 0.02& 8.65$\pm$ 0.01& 9.92$\pm$ 0.02&99.99$\pm$ 9.99&99.99$\pm$ 9.99&99.99$\pm$ 9.99&\\
ClassI0 11&18:02:12.69-22:46:07.0& 5.12$\pm$ 0.22& 9.62$\pm$ 0.13&10.11$\pm$ 0.08&10.85$\pm$ 0.06&11.55$\pm$ 0.09&99.99$\pm$ 9.99&99.99$\pm$ 9.99&99.99$\pm$ 9.99&\\
ClassI0 12&18:02:12.77-23:05:46.7& 1.83$\pm$ 0.02& 5.92$\pm$ 0.01& 6.77$\pm$ 0.01& 8.25$\pm$ 0.01& 9.87$\pm$ 0.02&14.23$\pm$ 0.02&15.61$\pm$ 0.06&17.17$\pm$ 0.04&\\
ClassI0 13&18:02:13.08-23:06:07.2& 3.89$\pm$ 0.11& 8.59$\pm$ 0.10& 9.19$\pm$ 0.04& 9.75$\pm$ 0.03&11.12$\pm$ 0.04&15.46$\pm$ 0.05&99.99$\pm$ 9.99&99.99$\pm$ 9.99&\\
ClassI0 14&18:02:14.04-23:06:40.7& 3.47$\pm$ 0.09& 8.16$\pm$ 0.11& 8.74$\pm$ 0.06&10.15$\pm$ 0.04&11.69$\pm$ 0.08&16.71$\pm$ 0.15&99.99$\pm$ 9.99&19.53$\pm$ 0.31&\\
ClassI0 15&18:02:15.79-23:06:42.8& 3.57$\pm$ 0.10& 7.51$\pm$ 0.06& 7.47$\pm$ 0.01& 8.26$\pm$ 0.01& 9.45$\pm$ 0.02&15.06$\pm$ 0.02&18.12$\pm$ 0.44&99.99$\pm$ 9.99&\\
ClassI0 16&18:02:16.80-23:03:47.2& 1.94$\pm$ 0.08& 7.61$\pm$ 0.04& 8.91$\pm$ 0.04& 9.58$\pm$ 0.03&10.53$\pm$ 0.04&11.93$\pm$ 0.02&12.26$\pm$ 0.02&12.99$\pm$ 0.02&\\
ClassI0 17&18:02:16.85-23:00:51.8& 1.13$\pm$ 0.06& 5.39$\pm$ 0.02& 6.12$\pm$ 0.01& 7.25$\pm$ 0.01& 8.38$\pm$ 0.01&14.61$\pm$ 0.02&17.53$\pm$ 0.21&99.99$\pm$ 9.99&\\
ClassI0 18&18:02:22.92-22:55:41.2& 0.81$\pm$ 0.01& 3.57$\pm$ 0.00& 4.45$\pm$ 0.00& 6.02$\pm$ 0.00& 7.32$\pm$ 0.01&12.44$\pm$ 0.02&14.81$\pm$ 0.04&15.62$\pm$ 0.02&\\
ClassI0 19&18:02:24.68-23:01:17.7& 2.02$\pm$ 0.90& 8.87$\pm$ 0.55& 9.19$\pm$ 0.17& 9.70$\pm$ 0.06&10.68$\pm$ 0.11&17.24$\pm$ 0.35&17.91$\pm$ 0.29&19.43$\pm$ 0.27&\\
ClassI0 20&18:02:25.25-22:46:04.4& 3.05$\pm$ 0.04& 7.78$\pm$ 0.06& 8.61$\pm$ 0.03& 9.41$\pm$ 0.02&10.21$\pm$ 0.03&13.68$\pm$ 0.02&15.86$\pm$ 0.08&17.98$\pm$ 0.08&\\
ClassI0 21&18:02:25.90-22:46:00.1& 2.34$\pm$ 0.02& 7.10$\pm$ 0.03& 9.00$\pm$ 0.05&99.99$\pm$ 9.99&11.51$\pm$ 0.08&13.59$\pm$ 0.02&14.52$\pm$ 0.02&16.18$\pm$ 0.02&\\
ClassI0 22&18:02:25.99-22:45:51.1& 2.53$\pm$ 0.03& 6.30$\pm$ 0.01& 7.25$\pm$ 0.01& 7.96$\pm$ 0.01& 8.66$\pm$ 0.01&11.23$\pm$ 0.02&13.25$\pm$ 0.02&16.11$\pm$ 0.02&\\
ClassI0 23&18:02:27.26-23:03:20.5& 1.75$\pm$ 0.08& 5.74$\pm$ 0.01& 6.35$\pm$ 0.01& 7.45$\pm$ 0.01& 8.86$\pm$ 0.02&11.78$\pm$ 0.02&15.27$\pm$ 0.04&99.99$\pm$ 9.99&\\
ClassI0 24&18:02:28.49-23:03:56.9& 2.09$\pm$ 0.11& 5.76$\pm$ 0.02& 7.69$\pm$ 0.03&10.48$\pm$ 0.06&10.78$\pm$ 0.07&13.57$\pm$ 0.02&16.20$\pm$ 0.08&19.15$\pm$ 0.20&\\
ClassI0 25&18:02:30.34-23:00:22.7& 1.82$\pm$ 0.04& 7.46$\pm$ 0.13& 8.40$\pm$ 0.07& 9.56$\pm$ 0.03&10.80$\pm$ 0.06&16.17$\pm$ 0.03&17.58$\pm$ 0.27&99.99$\pm$ 9.99&\\
ClassI0 26&18:02:31.71-22:56:47.0& 1.58$\pm$ 0.01& 4.80$\pm$ 0.01& 5.45$\pm$ 0.00& 6.29$\pm$ 0.00& 6.92$\pm$ 0.01& 9.07$\pm$ 0.02&11.34$\pm$ 0.02&15.40$\pm$ 0.02&\\
ClassI0 27&18:02:34.04-23:06:56.9& 5.16$\pm$ 0.50& 8.07$\pm$ 0.09& 8.71$\pm$ 0.04& 9.48$\pm$ 0.02&10.10$\pm$ 0.04&12.69$\pm$ 0.02&14.41$\pm$ 0.03&18.15$\pm$ 0.10&\\
ClassI0 28&18:02:34.17-23:06:53.3& 5.16$\pm$ 0.50& 8.04$\pm$ 0.08& 8.70$\pm$ 0.04& 9.53$\pm$ 0.02&10.06$\pm$ 0.03&11.16$\pm$ 0.02&12.85$\pm$ 0.02&16.47$\pm$ 0.02&\\
ClassI0 29&18:02:34.46-23:08:06.4& 1.99$\pm$ 0.01& 4.73$\pm$ 0.01& 5.23$\pm$ 0.00& 6.07$\pm$ 0.00& 6.64$\pm$ 0.00& 9.66$\pm$ 0.02&12.38$\pm$ 0.02&16.98$\pm$ 0.03&\\
ClassI0 30&18:02:34.85-22:49:55.2& 2.07$\pm$ 0.04& 5.86$\pm$ 0.03& 6.68$\pm$ 0.01& 7.79$\pm$ 0.01& 8.74$\pm$ 0.01&12.53$\pm$ 0.02&16.24$\pm$ 0.10&18.78$\pm$ 0.13&\\
ClassI0 31&18:02:40.73-23:00:30.6& 1.75$\pm$ 0.01& 5.39$\pm$ 0.02& 6.01$\pm$ 0.01& 6.88$\pm$ 0.01& 7.42$\pm$ 0.01& 9.60$\pm$ 0.02&12.13$\pm$ 0.02&16.88$\pm$ 0.04&\\
ClassI0 32&18:02:41.30-22:44:38.4& 1.69$\pm$ 0.03& 5.84$\pm$ 0.03& 7.63$\pm$ 0.03&99.99$\pm$ 9.99&11.52$\pm$ 0.16&14.02$\pm$ 0.02&14.88$\pm$ 0.03&16.31$\pm$ 0.02&\\
ClassI0 33&18:02:42.41-22:44:12.8& 0.50$\pm$ 0.01& 3.97$\pm$ 0.00& 5.77$\pm$ 0.01& 8.75$\pm$ 0.03& 8.97$\pm$ 0.03&14.05$\pm$ 0.05&14.78$\pm$ 0.05&16.26$\pm$ 0.04&\\
ClassI0 34&18:02:42.41-22:44:12.8& 0.50$\pm$ 0.01& 3.97$\pm$ 0.00& 5.77$\pm$ 0.01& 8.75$\pm$ 0.03& 8.97$\pm$ 0.03&14.05$\pm$ 0.05&14.78$\pm$ 0.05&16.26$\pm$ 0.04&\\
ClassI0 35&18:02:47.26-23:06:13.0& 3.58$\pm$ 0.09& 7.44$\pm$ 0.08& 8.89$\pm$ 0.07&10.15$\pm$ 0.03&10.31$\pm$ 0.03&12.98$\pm$ 0.02&14.40$\pm$ 0.02&16.28$\pm$ 0.02&\\
ClassI0 36&18:02:56.33-22:43:58.1& 5.32$\pm$ 0.18& 8.15$\pm$ 0.05& 9.11$\pm$ 0.04&10.66$\pm$ 0.05&10.69$\pm$ 0.05&99.99$\pm$ 9.99&99.99$\pm$ 9.99&99.99$\pm$ 9.99&\\
ClassI0 37&18:02:56.78-22:50:31.6& 3.80$\pm$ 0.05& 7.25$\pm$ 0.02& 8.10$\pm$ 0.02& 9.04$\pm$ 0.02&10.11$\pm$ 0.04&99.99$\pm$ 9.99&99.99$\pm$ 9.99&99.99$\pm$ 9.99&\\
Hotex  1&18:01:53.42-22:51:55.4& 2.76$\pm$ 0.01& 4.54$\pm$ 0.00& 5.19$\pm$ 0.00& 6.37$\pm$ 0.00& 8.09$\pm$ 0.01&12.23$\pm$ 0.02&99.99$\pm$ 9.99&99.99$\pm$ 9.99&\\
Hotex  2&18:01:59.33-22:52:36.1& 1.10$\pm$ 0.01& 3.66$\pm$ 0.00& 4.28$\pm$ 0.00& 5.56$\pm$ 0.00& 6.32$\pm$ 0.00& 8.34$\pm$ 0.02&10.56$\pm$ 0.02&14.85$\pm$ 0.02&\\
Hotex  3&18:01:59.36-23:07:38.6& 1.15$\pm$ 0.00& 3.62$\pm$ 0.00& 4.25$\pm$ 0.00& 5.57$\pm$ 0.00& 6.75$\pm$ 0.01& 9.62$\pm$ 0.02&12.29$\pm$ 0.02&99.99$\pm$ 9.99&\\
Hotex  4&18:02:04.85-23:07:19.2& 1.35$\pm$ 0.01& 3.56$\pm$ 0.00& 3.85$\pm$ 0.00& 5.33$\pm$ 0.00& 6.14$\pm$ 0.00&10.52$\pm$ 0.02&17.00$\pm$ 0.19&99.99$\pm$ 9.99&\\
Hotex  5&18:02:05.52-23:06:51.1& 1.50$\pm$ 0.01& 4.03$\pm$ 0.00& 4.24$\pm$ 0.00& 5.55$\pm$ 0.00& 6.09$\pm$ 0.00& 8.34$\pm$ 0.02&12.08$\pm$ 0.02&99.99$\pm$ 9.99&\\
Hotex  6&18:02:05.52-23:04:39.4& 7.34$\pm$ 0.59& 9.59$\pm$ 0.01& 9.49$\pm$ 0.05&10.10$\pm$ 0.03&11.57$\pm$ 0.05&99.99 $\pm$ 9.99&20.07$\pm$ 0.78&20.11$\pm$ 0.42&\\
Hotex  7&18:02:08.95-22:45:29.5& 1.45$\pm$ 0.01& 4.07$\pm$ 0.00& 4.96$\pm$ 0.00&99.99$\pm$ 9.99& 6.99$\pm$ 0.01&99.99$\pm$ 9.99&99.99$\pm$ 9.99&99.99$\pm$ 9.99&\\
Hotex  8&18:02:12.48-23:05:16.1& 4.88$\pm$ 0.28& 6.74$\pm$ 0.01& 7.48$\pm$ 0.01& 8.37$\pm$ 0.01& 9.18$\pm$ 0.02&12.03$\pm$ 0.02&13.91$\pm$ 0.02&16.28$\pm$ 0.02&\\
Hotex  9&18:02:12.50-23:05:15.4& 4.88$\pm$ 0.28& 6.75$\pm$ 0.01& 7.47$\pm$ 0.01& 8.37$\pm$ 0.01& 9.18$\pm$ 0.02&12.03$\pm$ 0.02&13.91$\pm$ 0.02&16.28$\pm$ 0.02&\\
Hotex 10&18:02:17.64-22:56:54.2& 1.74$\pm$ 0.02& 3.86$\pm$ 0.00& 4.32$\pm$ 0.00& 5.45$\pm$ 0.00& 6.13$\pm$ 0.01& 8.32$\pm$ 0.02& 9.79$\pm$ 0.02&13.31$\pm$ 0.02&\\
Hotex 11&18:02:22.75-23:09:30.6& 1.65$\pm$ 0.01& 4.06$\pm$ 0.00& 4.48$\pm$ 0.00& 5.72$\pm$ 0.00& 6.20$\pm$ 0.00& 8.22$\pm$ 0.02&10.27$\pm$ 0.02&14.69$\pm$ 0.02&\\
Hotex 12&18:02:49.70-22:49:05.5& 3.25$\pm$ 0.02& 4.80$\pm$ 0.00& 5.11$\pm$ 0.00& 6.00$\pm$ 0.00& 6.80$\pm$ 0.01& 9.21$\pm$ 0.02&10.45$\pm$ 0.02&13.17$\pm$ 0.02&\\
Hotex 13&18:02:50.79-23:11:29.8& 2.60$\pm$ 0.01& 4.40$\pm$ 0.00& 5.29$\pm$ 0.00& 6.09$\pm$ 0.00& 7.75$\pm$ 0.01&11.68$\pm$ 0.02&15.64$\pm$ 0.08&99.99$\pm$ 9.99&\\
ClassII  1&18:01:51.67-22:49:49.8& 3.19$\pm$ 0.01& 5.29$\pm$ 0.01& 5.65$\pm$ 0.00& 6.06$\pm$ 0.00& 6.51$\pm$ 0.00&99.99$\pm$ 9.99&99.99$\pm$ 9.99&99.99$\pm$ 9.99&\\
ClassII  2&18:01:52.18-22:52:45.8& 3.51$\pm$ 0.03& 6.55$\pm$ 0.01& 6.62$\pm$ 0.01& 6.80$\pm$ 0.01& 6.88$\pm$ 0.01& 8.38$\pm$ 0.02& 9.52$\pm$ 0.02&99.99$\pm$ 9.99&\\
ClassII  3&18:01:52.54-22:50:06.4& 2.78$\pm$ 0.01& 4.85$\pm$ 0.00& 5.17$\pm$ 0.00& 5.78$\pm$ 0.00& 6.42$\pm$ 0.00&99.99$\pm$ 9.99&99.99$\pm$ 9.99&99.99$\pm$ 9.99&\\
ClassII  4&18:01:53.81-23:03:34.9& 4.57$\pm$ 0.03& 8.62$\pm$ 0.05& 8.96$\pm$ 0.03&99.99$\pm$ 9.99& 9.57$\pm$ 0.02&11.67$\pm$ 0.02&13.09$\pm$ 0.02&14.99$\pm$ 0.02&\\
ClassII  5&18:01:54.05-23:11:51.7& 3.91$\pm$ 0.04& 7.44$\pm$ 0.02& 8.31$\pm$ 0.02& 8.32$\pm$ 0.01& 8.96$\pm$ 0.01&11.82$\pm$ 0.02&15.12$\pm$ 0.04&21.13$\pm$ 0.75&\\
ClassII  6&18:01:54.46-23:05:51.7& 4.67$\pm$ 0.05& 7.24$\pm$ 0.01& 7.27$\pm$ 0.01&99.99$\pm$ 9.99& 7.85$\pm$ 0.01& 9.67$\pm$ 0.02&12.02$\pm$ 0.02&16.67$\pm$ 0.03&\\
ClassII  7&18:01:54.55-22:53:58.6& 3.27$\pm$ 0.03& 6.36$\pm$ 0.01& 6.39$\pm$ 0.01& 6.82$\pm$ 0.01& 6.87$\pm$ 0.01& 8.43$\pm$ 0.02& 9.86$\pm$ 0.02&13.07$\pm$ 0.02&\\
ClassII  8&18:01:55.06-22:49:45.5& 4.36$\pm$ 0.05& 7.52$\pm$ 0.03& 7.55$\pm$ 0.01& 8.01$\pm$ 0.01& 8.15$\pm$ 0.01&99.99$\pm$ 9.99&99.99$\pm$ 9.99&99.99$\pm$ 9.99&\\
ClassII  9&18:01:55.34-22:46:43.7& 3.82$\pm$ 0.02& 6.55$\pm$ 0.01& 7.07$\pm$ 0.01& 7.58$\pm$ 0.01& 7.99$\pm$ 0.01&99.99$\pm$ 9.99&99.99$\pm$ 9.99&99.99$\pm$ 9.99&\\
ClassII 10&18:01:55.73-22:55:16.7& 3.77$\pm$ 0.04& 6.03$\pm$ 0.01& 6.15$\pm$ 0.01& 6.43$\pm$ 0.00& 6.68$\pm$ 0.01& 8.35$\pm$ 0.02& 9.89$\pm$ 0.02&13.37$\pm$ 0.02&\\
ClassII 11&18:01:58.08-22:51:31.7& 4.44$\pm$ 0.09& 6.75$\pm$ 0.01& 6.84$\pm$ 0.01& 7.35$\pm$ 0.01& 7.25$\pm$ 0.01&99.99$\pm$ 9.99&99.99$\pm$ 9.99&99.99$\pm$ 9.99&\\
ClassII 12&18:01:58.61-22:44:49.6& 4.92$\pm$ 0.08& 7.47$\pm$ 0.04& 7.33$\pm$ 0.02&99.99$\pm$ 9.99& 7.45$\pm$ 0.01&99.99$\pm$ 9.99&99.99$\pm$ 9.99&99.99$\pm$ 9.99&\\
ClassII 13&18:01:58.63-23:01:35.0& 5.06$\pm$ 0.29& 7.39$\pm$ 0.03& 7.44$\pm$ 0.01& 7.85$\pm$ 0.01& 8.07$\pm$ 0.01& 9.79$\pm$ 0.02&11.40$\pm$ 0.02&14.83$\pm$ 0.02&\\
ClassII 14&18:01:58.65-23:06:19.8& 4.58$\pm$ 0.08& 8.32$\pm$ 0.04& 8.29$\pm$ 0.02& 8.69$\pm$ 0.01& 8.71$\pm$ 0.01&10.34$\pm$ 0.02&12.53$\pm$ 0.02&16.86$\pm$ 0.03&\\
ClassII 15&18:01:58.97-23:08:23.6& 4.52$\pm$ 0.05& 6.82$\pm$ 0.02& 6.76$\pm$ 0.01& 7.19$\pm$ 0.01& 7.33$\pm$ 0.01& 9.08$\pm$ 0.02&11.29$\pm$ 0.02&15.72$\pm$ 0.02&\\
ClassII 16&18:01:59.38-23:03:07.6& 3.37$\pm$ 0.05& 6.37$\pm$ 0.01& 6.65$\pm$ 0.01& 7.06$\pm$ 0.01& 7.58$\pm$ 0.01&10.17$\pm$ 0.02&12.77$\pm$ 0.02&99.99$\pm$ 9.99&\\
ClassII 17&18:01:59.38-22:48:12.6& 4.92$\pm$ 0.05& 8.05$\pm$ 0.04& 7.97$\pm$ 0.02& 8.41$\pm$ 0.01& 8.56$\pm$ 0.01&99.99$\pm$ 9.99&99.99$\pm$ 9.99&99.99$\pm$ 9.99&\\
ClassII 18&18:01:59.40-23:05:56.8& 4.23$\pm$ 0.05& 6.28$\pm$ 0.01& 6.38$\pm$ 0.01& 6.78$\pm$ 0.01& 7.14$\pm$ 0.01& 8.97$\pm$ 0.02&11.21$\pm$ 0.02&15.60$\pm$ 0.02&\\
ClassII 19&18:02:01.56-23:01:11.3& 5.18$\pm$ 0.38& 8.12$\pm$ 0.07& 8.13$\pm$ 0.02& 8.41$\pm$ 0.01& 8.78$\pm$ 0.01&10.22$\pm$ 0.02&12.32$\pm$ 0.02&16.52$\pm$ 0.02&\\
ClassII 20&18:02:02.31-22:49:12.7& 4.47$\pm$ 0.05& 8.67$\pm$ 0.04& 8.65$\pm$ 0.02& 9.16$\pm$ 0.02& 9.20$\pm$ 0.02&99.99$\pm$ 9.99&99.99$\pm$ 9.99&99.99$\pm$ 9.99&\\
ClassII 21&18:02:03.19-22:48:48.6& 1.75$\pm$ 0.01& 5.07$\pm$ 0.00& 5.69$\pm$ 0.00& 6.26$\pm$ 0.00& 6.60$\pm$ 0.00&99.99$\pm$ 9.99&99.99$\pm$ 9.99&99.99$\pm$ 9.99&\\
ClassII 22&18:02:05.62-22:46:39.7& 4.91$\pm$ 0.10& 7.70$\pm$ 0.03& 7.60$\pm$ 0.01& 8.06$\pm$ 0.01& 8.04$\pm$ 0.01&99.99$\pm$ 9.99&99.99$\pm$ 9.99&99.99$\pm$ 9.99&\\
ClassII 23&18:02:05.86-22:47:41.6& 5.25$\pm$ 0.20& 8.13$\pm$ 0.03& 8.08$\pm$ 0.02& 8.49$\pm$ 0.01& 8.85$\pm$ 0.01&99.99$\pm$ 9.99&99.99$\pm$ 9.99&99.99$\pm$ 9.99&\\
ClassII 24&18:02:07.15-22:49:07.0& 2.72$\pm$ 0.02& 6.96$\pm$ 0.02& 6.96$\pm$ 0.01& 7.28$\pm$ 0.01& 7.41$\pm$ 0.01&99.99$\pm$ 9.99&99.99$\pm$ 9.99&99.99$\pm$ 9.99&\\
ClassII 25&18:02:08.38-22:49:04.4& 3.70$\pm$ 0.03& 7.00$\pm$ 0.01& 7.15$\pm$ 0.01& 7.75$\pm$ 0.01& 7.92$\pm$ 0.01&99.99$\pm$ 9.99&99.99$\pm$ 9.99&99.99$\pm$ 9.99&\\
ClassII 26&18:02:08.59-22:48:59.0& 3.70$\pm$ 0.03& 6.55$\pm$ 0.01& 6.66$\pm$ 0.01& 6.72$\pm$ 0.01& 6.78$\pm$ 0.01&99.99$\pm$ 9.99&99.99$\pm$ 9.99&99.99$\pm$ 9.99&\\
ClassII 27&18:02:09.60-22:49:28.9& 4.20$\pm$ 0.14& 7.39$\pm$ 0.02& 7.29$\pm$ 0.01& 7.82$\pm$ 0.01& 7.63$\pm$ 0.01&99.99$\pm$ 9.99&99.99$\pm$ 9.99&99.99$\pm$ 9.99&\\
ClassII 28&18:02:09.74-22:56:01.3& 2.60$\pm$ 0.03& 5.61$\pm$ 0.01& 5.79$\pm$ 0.01& 6.24$\pm$ 0.00& 6.71$\pm$ 0.00& 8.36$\pm$ 0.02&10.27$\pm$ 0.02&14.54$\pm$ 0.02&\\
ClassII 29&18:02:10.30-22:54:55.8& 4.36$\pm$ 0.14& 7.30$\pm$ 0.03& 7.75$\pm$ 0.02& 8.37$\pm$ 0.01& 8.73$\pm$ 0.01&11.25$\pm$ 0.02&16.47$\pm$ 0.02&99.99$\pm$ 9.99&\\
ClassII 30&18:02:10.94-22:47:56.8& 4.67$\pm$ 0.16& 7.76$\pm$ 0.04& 7.79$\pm$ 0.02& 8.20$\pm$ 0.01& 8.40$\pm$ 0.01&99.99$\pm$ 9.99&99.99$\pm$ 9.99&99.99$\pm$ 9.99&\\
ClassII 31&18:02:11.71-22:47:42.4& 5.68$\pm$ 0.45& 9.06$\pm$ 0.09& 9.78$\pm$ 0.07&10.34$\pm$ 0.04&10.68$\pm$ 0.04&99.99$\pm$ 9.99&99.99$\pm$ 9.99&99.99$\pm$ 9.99&\\
ClassII 32&18:02:12.67-22:58:52.7& 2.87$\pm$ 0.08& 6.16$\pm$ 0.02& 6.19$\pm$ 0.01& 6.50$\pm$ 0.00& 6.60$\pm$ 0.00& 8.22$\pm$ 0.02& 9.70$\pm$ 0.02&13.18$\pm$ 0.02&\\
ClassII 33&18:02:12.89-22:54:40.7& 4.62$\pm$ 0.17& 6.80$\pm$ 0.02& 6.85$\pm$ 0.01& 7.00$\pm$ 0.01& 6.92$\pm$ 0.01& 7.93$\pm$ 0.02& 8.76$\pm$ 0.02&10.54$\pm$ 0.02&\\
ClassII 34&18:02:12.99-22:49:45.5& 4.70$\pm$ 0.09& 7.74$\pm$ 0.03& 7.82$\pm$ 0.02& 7.96$\pm$ 0.01& 7.84$\pm$ 0.01&99.99$\pm$ 9.99&99.99$\pm$ 9.99&99.99$\pm$ 9.99&\\
ClassII 35&18:02:14.21-23:01:44.0& 0.75$\pm$ 0.04& 3.93$\pm$ 0.00& 5.26$\pm$ 0.00& 5.82$\pm$ 0.00& 6.57$\pm$ 0.00& 8.74$\pm$ 0.02&10.21$\pm$ 0.02&11.99$\pm$ 0.02&\\
ClassII 36&18:02:14.33-22:45:57.2& 4.18$\pm$ 0.05& 7.27$\pm$ 0.02& 7.24$\pm$ 0.01& 7.66$\pm$ 0.01& 7.80$\pm$ 0.01&99.99$\pm$ 9.99&99.99$\pm$ 9.99&99.99$\pm$ 9.99&\\
ClassII 37&18:02:15.86-23:05:55.7& 3.68$\pm$ 0.14& 7.64$\pm$ 0.05& 9.32$\pm$ 0.06&10.15$\pm$ 0.04&10.47$\pm$ 0.03&11.72$\pm$ 0.02&12.33$\pm$ 0.02&13.39$\pm$ 0.02&\\
ClassII 38&18:02:16.22-22:48:33.1& 4.28$\pm$ 0.07& 6.48$\pm$ 0.01& 6.54$\pm$ 0.01& 6.76$\pm$ 0.01& 7.05$\pm$ 0.01& 8.41$\pm$ 0.02& 9.53$\pm$ 0.02&99.99$\pm$ 9.99&\\
ClassII 39&18:02:16.77-22:54:37.4& 3.75$\pm$ 0.08& 6.33$\pm$ 0.01& 6.38$\pm$ 0.01& 6.71$\pm$ 0.01& 6.94$\pm$ 0.01& 8.75$\pm$ 0.02&11.28$\pm$ 0.02&16.23$\pm$ 0.02&\\
ClassII 40&18:02:16.97-23:09:46.8& 4.81$\pm$ 0.06& 7.45$\pm$ 0.03& 7.37$\pm$ 0.01& 7.77$\pm$ 0.01& 7.90$\pm$ 0.01& 9.40$\pm$ 0.02&11.38$\pm$ 0.02&15.35$\pm$ 0.02&\\
ClassII 41&18:02:17.69-23:07:17.0& 4.32$\pm$ 0.25& 8.06$\pm$ 0.10& 8.07$\pm$ 0.03& 8.76$\pm$ 0.02& 9.12$\pm$ 0.02&12.33$\pm$ 0.02&16.20$\pm$ 0.11&99.99$\pm$ 9.99&\\
ClassII 42&18:02:18.19-23:07:50.5& 3.68$\pm$ 0.03& 6.48$\pm$ 0.04& 6.74$\pm$ 0.02& 7.27$\pm$ 0.01& 7.64$\pm$ 0.01& 9.99$\pm$ 0.02&12.44$\pm$ 0.02&17.29$\pm$ 0.04&\\
ClassII 43&18:02:18.60-22:49:48.0& 4.25$\pm$ 0.07& 6.85$\pm$ 0.01& 6.89$\pm$ 0.01& 7.23$\pm$ 0.01& 7.22$\pm$ 0.01& 8.91$\pm$ 0.02&10.84$\pm$ 0.02&99.99$\pm$ 9.99&\\
ClassII 44&18:02:19.25-22:55:13.1& 3.66$\pm$ 0.02& 8.31$\pm$ 0.10& 9.37$\pm$ 0.09& 9.90$\pm$ 0.03& 9.86$\pm$ 0.03&10.71$\pm$ 0.02&11.84$\pm$ 0.02&14.23$\pm$ 0.02&\\
ClassII 45&18:02:21.91-22:47:53.5& 5.44$\pm$ 0.10& 7.79$\pm$ 0.03& 7.81$\pm$ 0.02& 8.27$\pm$ 0.01& 8.20$\pm$ 0.01& 9.58$\pm$ 0.02&11.25$\pm$ 0.02&14.56$\pm$ 0.02&\\
ClassII 46&18:02:23.23-23:01:35.0& 2.65$\pm$ 0.30& 6.17$\pm$ 0.06& 6.56$\pm$ 0.03& 6.96$\pm$ 0.01& 7.34$\pm$ 0.01& 8.86$\pm$ 0.02& 9.95$\pm$ 0.02&11.31$\pm$ 0.02&\\
ClassII 47&18:02:25.37-23:06:15.8& 2.57$\pm$ 0.09& 5.70$\pm$ 0.01& 5.99$\pm$ 0.01& 6.66$\pm$ 0.01& 7.04$\pm$ 0.01& 9.25$\pm$ 0.02&12.15$\pm$ 0.02&17.38$\pm$ 0.04&\\
ClassII 48&18:02:25.47-22:50:19.7& 6.01$\pm$ 0.44& 8.03$\pm$ 0.03& 7.99$\pm$ 0.01& 8.26$\pm$ 0.01& 8.19$\pm$ 0.01& 9.32$\pm$ 0.02&10.77$\pm$ 0.02&13.72$\pm$ 0.02&\\
ClassII 49&18:02:26.93-22:50:37.3& 3.66$\pm$ 0.03& 6.19$\pm$ 0.01& 6.32$\pm$ 0.01& 6.67$\pm$ 0.01& 6.90$\pm$ 0.01& 8.24$\pm$ 0.02& 9.48$\pm$ 0.02&11.95$\pm$ 0.02&\\
ClassII 50&18:02:30.67-22:58:01.2& 3.18$\pm$ 0.05& 6.13$\pm$ 0.03& 6.19$\pm$ 0.01& 6.65$\pm$ 0.01& 6.98$\pm$ 0.01& 8.42$\pm$ 0.02&10.04$\pm$ 0.02&13.79$\pm$ 0.02&\\
ClassII 51&18:02:30.96-23:02:35.2& 2.91$\pm$ 0.14& 7.86$\pm$ 0.07& 8.88$\pm$ 0.04& 9.41$\pm$ 0.02& 9.64$\pm$ 0.02&10.91$\pm$ 0.02&11.56$\pm$ 0.02&12.15$\pm$ 0.02&\\
ClassII 52&18:02:31.77-22:57:53.6& 3.54$\pm$ 0.14& 6.07$\pm$ 0.03& 6.51$\pm$ 0.01& 6.62$\pm$ 0.01& 6.61$\pm$ 0.00& 7.89$\pm$ 0.02& 8.71$\pm$ 0.02&10.61$\pm$ 0.02&\\
ClassII 53&18:02:32.76-22:49:00.5& 5.41$\pm$ 0.16& 7.64$\pm$ 0.02& 7.72$\pm$ 0.01& 8.06$\pm$ 0.01& 7.89$\pm$ 0.01& 9.32$\pm$ 0.02&10.64$\pm$ 0.02&12.06$\pm$ 0.02&\\
ClassII 54&18:02:32.97-22:53:49.2& 4.10$\pm$ 0.04& 6.59$\pm$ 0.02& 6.71$\pm$ 0.01& 7.00$\pm$ 0.01& 7.06$\pm$ 0.01& 8.52$\pm$ 0.02& 9.95$\pm$ 0.02&13.23$\pm$ 0.02&\\
ClassII 55&18:02:33.27-23:02:23.6& 2.12$\pm$ 0.08& 5.55$\pm$ 0.01& 5.57$\pm$ 0.00& 6.03$\pm$ 0.00& 6.43$\pm$ 0.00& 8.29$\pm$ 0.02& 9.95$\pm$ 0.02&13.85$\pm$ 0.02&\\
ClassII 56&18:02:33.65-23:09:02.9& 3.71$\pm$ 0.02& 7.32$\pm$ 0.03& 7.37$\pm$ 0.01& 7.94$\pm$ 0.01& 7.83$\pm$ 0.01& 9.41$\pm$ 0.02&11.41$\pm$ 0.02&15.36$\pm$ 0.02&\\
ClassII 57&18:02:33.65-23:09:02.9& 3.71$\pm$ 0.02& 7.32$\pm$ 0.03& 7.37$\pm$ 0.01& 7.94$\pm$ 0.01& 7.83$\pm$ 0.01& 9.41$\pm$ 0.02&11.41$\pm$ 0.02&15.36$\pm$ 0.02&\\
ClassII 58&18:02:33.82-22:53:26.9& 4.19$\pm$ 0.04& 6.83$\pm$ 0.01& 6.84$\pm$ 0.01& 7.23$\pm$ 0.01& 7.18$\pm$ 0.01& 8.78$\pm$ 0.02&10.53$\pm$ 0.02&99.99$\pm$ 9.99&\\
ClassII 59&18:02:34.44-22:59:16.4& 4.17$\pm$ 0.22& 6.68$\pm$ 0.02& 7.30$\pm$ 0.01& 7.99$\pm$ 0.01& 8.42$\pm$ 0.01&10.59$\pm$ 0.02&12.46$\pm$ 0.02&16.15$\pm$ 0.02&\\
ClassII 60&18:02:35.04-22:58:22.4& 2.99$\pm$ 0.06& 5.30$\pm$ 0.01& 5.33$\pm$ 0.00& 5.91$\pm$ 0.00& 6.02$\pm$ 0.00& 7.84$\pm$ 0.02& 8.61$\pm$ 0.02&10.52$\pm$ 0.02&\\
ClassII 61&18:02:35.23-22:58:10.6& 2.95$\pm$ 0.07& 7.54$\pm$ 0.10& 8.28$\pm$ 0.05& 9.14$\pm$ 0.02& 9.11$\pm$ 0.03&10.65$\pm$ 0.02&12.52$\pm$ 0.02&16.46$\pm$ 0.03&\\
ClassII 62&18:02:36.17-23:09:14.4& 4.83$\pm$ 0.10& 7.69$\pm$ 0.04& 7.73$\pm$ 0.01& 8.15$\pm$ 0.01& 8.26$\pm$ 0.01& 9.82$\pm$ 0.02&11.93$\pm$ 0.02&16.17$\pm$ 0.02&\\
ClassII 63&18:02:36.50-22:44:08.5& 5.05$\pm$ 0.32& 8.16$\pm$ 0.03& 8.08$\pm$ 0.02&99.99$\pm$ 9.99& 8.32$\pm$ 0.01& 9.29$\pm$ 0.02&10.62$\pm$ 0.02&13.59$\pm$ 0.02&\\
ClassII 64&18:02:37.35-22:48:57.2& 3.77$\pm$ 0.05& 7.39$\pm$ 0.03& 7.57$\pm$ 0.01& 7.88$\pm$ 0.01& 7.70$\pm$ 0.01& 9.13$\pm$ 0.02&10.47$\pm$ 0.02&13.31$\pm$ 0.02&\\
ClassII 65&18:02:37.65-22:46:59.5& 3.42$\pm$ 0.03& 6.49$\pm$ 0.02& 6.99$\pm$ 0.01& 7.25$\pm$ 0.01& 7.39$\pm$ 0.01& 8.55$\pm$ 0.02& 9.77$\pm$ 0.02&12.58$\pm$ 0.02&\\
ClassII 66&18:02:38.95-22:58:41.9& 3.73$\pm$ 0.09& 7.11$\pm$ 0.05& 7.29$\pm$ 0.01& 7.79$\pm$ 0.01& 8.16$\pm$ 0.01& 9.74$\pm$ 0.02&11.88$\pm$ 0.02&99.99$\pm$ 9.99&\\
ClassII 67&18:02:40.03-23:05:55.7& 3.52$\pm$ 0.08& 7.06$\pm$ 0.07& 7.02$\pm$ 0.02& 7.42$\pm$ 0.01& 7.51$\pm$ 0.01& 8.89$\pm$ 0.02&10.41$\pm$ 0.02&13.66$\pm$ 0.02&\\
ClassII 68&18:02:40.42-23:08:50.6& 4.93$\pm$ 0.08& 7.05$\pm$ 0.03& 7.35$\pm$ 0.01& 7.57$\pm$ 0.01& 7.84$\pm$ 0.01& 9.44$\pm$ 0.02&11.02$\pm$ 0.02&14.49$\pm$ 0.02&\\
ClassII 69&18:02:41.07-23:08:43.4& 4.78$\pm$ 0.07& 6.91$\pm$ 0.02& 7.28$\pm$ 0.01& 7.58$\pm$ 0.01& 7.73$\pm$ 0.01& 8.97$\pm$ 0.02&10.50$\pm$ 0.02&13.83$\pm$ 0.02&\\
ClassII 70&18:02:41.45-22:51:42.1& 3.07$\pm$ 0.02& 7.69$\pm$ 0.07& 7.76$\pm$ 0.02& 8.07$\pm$ 0.01& 8.23$\pm$ 0.01& 9.81$\pm$ 0.02&11.78$\pm$ 0.02&99.99$\pm$ 9.99&\\
ClassII 71&18:02:41.64-23:00:03.6& 2.95$\pm$ 0.05& 6.46$\pm$ 0.03& 6.65$\pm$ 0.01& 7.14$\pm$ 0.01& 7.39$\pm$ 0.01& 9.91$\pm$ 0.02&12.06$\pm$ 0.02&16.73$\pm$ 0.03&\\
ClassII 72&18:02:42.17-23:02:39.1& 3.00$\pm$ 0.06& 6.97$\pm$ 0.05& 7.56$\pm$ 0.02& 8.09$\pm$ 0.01& 8.33$\pm$ 0.01&10.12$\pm$ 0.02&12.26$\pm$ 0.02&15.99$\pm$ 0.02&\\
ClassII 73&18:02:42.99-23:03:13.0& 3.63$\pm$ 0.14& 7.53$\pm$ 0.04& 8.99$\pm$ 0.05&10.15$\pm$ 0.03&10.27$\pm$ 0.04&11.37$\pm$ 0.02&12.06$\pm$ 0.02&13.26$\pm$ 0.02&\\
ClassII 74&18:02:43.10-23:07:21.4& 3.72$\pm$ 0.06& 8.51$\pm$ 0.10& 8.53$\pm$ 0.03& 8.96$\pm$ 0.02& 9.02$\pm$ 0.01&10.56$\pm$ 0.02&12.70$\pm$ 0.02&16.82$\pm$ 0.03&\\
ClassII 75&18:02:43.82-23:05:30.1& 4.63$\pm$ 0.18& 6.92$\pm$ 0.02& 7.02$\pm$ 0.01& 7.48$\pm$ 0.01& 7.13$\pm$ 0.01& 8.51$\pm$ 0.02& 9.88$\pm$ 0.02&12.72$\pm$ 0.02&\\
ClassII 76&18:02:45.00-23:08:05.3& 4.29$\pm$ 0.06& 7.49$\pm$ 0.03& 7.52$\pm$ 0.01& 7.97$\pm$ 0.01& 7.86$\pm$ 0.01& 9.48$\pm$ 0.02&11.28$\pm$ 0.02&14.83$\pm$ 0.02&\\
ClassII 77&18:02:45.07-23:03:17.3& 4.36$\pm$ 0.20& 7.70$\pm$ 0.05& 7.86$\pm$ 0.02& 8.23$\pm$ 0.01& 8.34$\pm$ 0.01& 9.77$\pm$ 0.02&11.46$\pm$ 0.02&15.06$\pm$ 0.02&\\
ClassII 78&18:02:45.38-22:48:51.8& 3.39$\pm$ 0.05& 6.67$\pm$ 0.01& 6.59$\pm$ 0.01& 7.16$\pm$ 0.01& 7.36$\pm$ 0.01& 8.94$\pm$ 0.02&10.72$\pm$ 0.02&14.34$\pm$ 0.02&\\
ClassII 79&18:02:45.41-22:46:07.0& 3.80$\pm$ 0.05& 7.29$\pm$ 0.03& 8.42$\pm$ 0.04& 9.17$\pm$ 0.02& 9.06$\pm$ 0.02& 9.16$\pm$ 0.02& 9.44$\pm$ 0.02&10.62$\pm$ 0.02&\\
ClassII 80&18:02:45.81-22:44:22.9& 3.00$\pm$ 0.03& 7.90$\pm$ 0.09& 7.94$\pm$ 0.03&99.99$\pm$ 9.99& 8.05$\pm$ 0.01& 9.04$\pm$ 0.02&10.35$\pm$ 0.02&13.23$\pm$ 0.02&\\
ClassII 81&18:02:48.93-22:45:41.4& 4.47$\pm$ 0.14& 7.22$\pm$ 0.02& 7.69$\pm$ 0.01&99.99$\pm$ 9.99& 7.86$\pm$ 0.01& 9.82$\pm$ 0.02&12.01$\pm$ 0.02&16.21$\pm$ 0.02&\\
ClassII 82&18:02:49.20-22:59:45.6& 4.18$\pm$ 0.19& 8.86$\pm$ 0.15& 9.99$\pm$ 0.09&10.19$\pm$ 0.03&10.46$\pm$ 0.03&11.30$\pm$ 0.02&12.02$\pm$ 0.02&13.12$\pm$ 0.02&\\
ClassII 83&18:02:50.16-23:03:39.2& 3.08$\pm$ 0.03& 5.17$\pm$ 0.01& 5.62$\pm$ 0.00& 6.23$\pm$ 0.00& 6.72$\pm$ 0.01& 8.71$\pm$ 0.02& 9.99$\pm$ 0.02&12.97$\pm$ 0.02&\\
ClassII 84&18:02:50.45-22:48:50.0& 4.75$\pm$ 0.10& 7.40$\pm$ 0.02& 7.30$\pm$ 0.01& 7.40$\pm$ 0.01& 7.38$\pm$ 0.01& 8.01$\pm$ 0.02& 8.89$\pm$ 0.02&10.52$\pm$ 0.02&\\
ClassII 85&18:02:50.54-23:09:23.4& 4.55$\pm$ 0.03& 7.55$\pm$ 0.02& 8.08$\pm$ 0.01& 8.59$\pm$ 0.01& 8.41$\pm$ 0.01& 9.47$\pm$ 0.02&10.85$\pm$ 0.02&13.65$\pm$ 0.02&\\
ClassII 86&18:02:51.53-22:46:41.5& 4.80$\pm$ 0.14& 7.63$\pm$ 0.03& 7.62$\pm$ 0.01& 7.90$\pm$ 0.01& 7.81$\pm$ 0.01& 8.72$\pm$ 0.02&99.99$\pm$ 9.99&12.38$\pm$ 0.02&\\
ClassII 87&18:02:51.62-22:47:50.6& 4.61$\pm$ 0.10& 7.92$\pm$ 0.04& 8.41$\pm$ 0.02& 8.74$\pm$ 0.01& 8.59$\pm$ 0.01& 9.65$\pm$ 0.02&99.99$\pm$ 9.99&13.28$\pm$ 0.02&\\
ClassII 88&18:02:51.67-23:02:07.1& 4.93$\pm$ 0.16& 7.70$\pm$ 0.02& 7.69$\pm$ 0.01& 8.02$\pm$ 0.01& 7.95$\pm$ 0.01& 9.00$\pm$ 0.02&10.45$\pm$ 0.02&13.51$\pm$ 0.02&\\
ClassII 89&18:02:51.69-23:05:13.6& 4.74$\pm$ 0.06& 7.04$\pm$ 0.02& 7.23$\pm$ 0.01& 7.85$\pm$ 0.01& 7.73$\pm$ 0.01& 9.03$\pm$ 0.02&10.39$\pm$ 0.02&13.22$\pm$ 0.02&\\
ClassII 90&18:02:51.99-23:03:56.5& 3.85$\pm$ 0.06& 8.58$\pm$ 0.09&10.56$\pm$ 0.16&10.59$\pm$ 0.04&11.08$\pm$ 0.07&12.43$\pm$ 0.02&13.79$\pm$ 0.02&16.49$\pm$ 0.02&\\
ClassII 91&18:02:52.39-23:00:42.1& 3.73$\pm$ 0.09& 7.06$\pm$ 0.03& 7.04$\pm$ 0.01& 7.42$\pm$ 0.01& 7.51$\pm$ 0.01& 8.85$\pm$ 0.02&10.65$\pm$ 0.02&14.51$\pm$ 0.02&\\
ClassII 92&18:02:53.59-22:53:30.5& 5.56$\pm$ 0.07& 7.87$\pm$ 0.03& 7.87$\pm$ 0.01& 8.10$\pm$ 0.01& 8.09$\pm$ 0.01& 8.92$\pm$ 0.02&10.23$\pm$ 0.02&12.94$\pm$ 0.02&\\
ClassII 93&18:02:53.61-22:45:32.0& 4.72$\pm$ 0.11& 8.07$\pm$ 0.04& 8.26$\pm$ 0.02&99.99$\pm$ 9.99& 8.47$\pm$ 0.01& 9.55$\pm$ 0.02&99.99$\pm$ 9.99&13.90$\pm$ 0.02&\\
ClassII 94&18:02:53.98-23:09:22.3& 5.11$\pm$ 0.07& 7.25$\pm$ 0.02& 7.46$\pm$ 0.01& 7.59$\pm$ 0.01& 7.61$\pm$ 0.01& 8.68$\pm$ 0.02& 9.83$\pm$ 0.02&12.34$\pm$ 0.02&\\
ClassII 95&18:02:55.42-22:46:34.3& 4.30$\pm$ 0.05& 6.91$\pm$ 0.02& 7.16$\pm$ 0.01& 7.77$\pm$ 0.01& 8.06$\pm$ 0.01&99.99$\pm$ 9.99&99.99$\pm$ 9.99&99.99$\pm$ 9.99&\\
ClassII 96&18:02:56.18-23:05:10.0& 3.72$\pm$ 0.05& 6.92$\pm$ 0.04& 7.20$\pm$ 0.01& 7.54$\pm$ 0.01& 7.68$\pm$ 0.01& 8.87$\pm$ 0.02&10.24$\pm$ 0.02&13.23$\pm$ 0.02&\\
ClassII 97&18:02:57.29-22:46:11.6& 5.46$\pm$ 0.11& 7.84$\pm$ 0.05& 7.79$\pm$ 0.02& 8.12$\pm$ 0.01& 8.29$\pm$ 0.01&99.99$\pm$ 9.99&99.99$\pm$ 9.99&99.99$\pm$ 9.99&\\
ClassII 98&18:02:57.46-22:57:26.6& 4.78$\pm$ 0.04& 8.82$\pm$ 0.12& 8.50$\pm$ 0.03& 8.73$\pm$ 0.02& 8.90$\pm$ 0.01&10.31$\pm$ 0.02&12.33$\pm$ 0.02&16.22$\pm$ 0.02&\\
ClassII 99&18:02:57.55-22:47:29.0& 5.33$\pm$ 0.05& 7.66$\pm$ 0.02& 7.74$\pm$ 0.01& 8.15$\pm$ 0.01& 7.99$\pm$ 0.01&99.99$\pm$ 9.99&99.99$\pm$ 9.99&99.99$\pm$ 9.99&\\
ClassII100&18:02:57.60-22:45:28.4& 4.86$\pm$ 0.08& 7.85$\pm$ 0.05& 8.21$\pm$ 0.02&99.99$\pm$ 9.99& 8.91$\pm$ 0.02&99.99$\pm$ 9.99&99.99$\pm$ 9.99&99.99$\pm$ 9.99&\\
ClassII101&18:02:58.03-22:51:57.6& 4.49$\pm$ 0.05& 7.29$\pm$ 0.02& 7.33$\pm$ 0.01& 7.59$\pm$ 0.01& 7.44$\pm$ 0.01& 8.31$\pm$ 0.02& 9.18$\pm$ 0.02&10.85$\pm$ 0.02&\\
ClassII102&18:02:58.27-22:59:53.5& 4.65$\pm$ 0.07& 6.92$\pm$ 0.02& 6.97$\pm$ 0.01& 7.09$\pm$ 0.01& 7.21$\pm$ 0.01& 8.02$\pm$ 0.02& 8.94$\pm$ 0.02&10.64$\pm$ 0.02&\\
ClassII103&18:03:00.17-23:00:34.2& 5.39$\pm$ 0.10& 8.31$\pm$ 0.03& 8.58$\pm$ 0.02& 8.89$\pm$ 0.02& 8.93$\pm$ 0.01&10.30$\pm$ 0.02&11.91$\pm$ 0.02&15.34$\pm$ 0.02&\\
ClassII104&18:03:00.19-23:03:07.6& 5.23$\pm$ 0.15& 7.45$\pm$ 0.03& 7.43$\pm$ 0.01& 7.69$\pm$ 0.01& 7.70$\pm$ 0.01& 8.73$\pm$ 0.02& 9.96$\pm$ 0.02&12.63$\pm$ 0.02&\\
ClassII105&18:03:01.20-23:03:35.3& 3.82$\pm$ 0.04& 7.38$\pm$ 0.03& 7.58$\pm$ 0.01& 7.83$\pm$ 0.01& 7.83$\pm$ 0.01& 8.90$\pm$ 0.02&10.27$\pm$ 0.02&99.99$\pm$ 9.99&\\
ClassII106&18:03:02.14-22:59:48.5& 4.65$\pm$ 0.07& 6.97$\pm$ 0.02& 7.22$\pm$ 0.01& 7.62$\pm$ 0.01& 7.40$\pm$ 0.01& 8.25$\pm$ 0.02& 9.28$\pm$ 0.02&11.07$\pm$ 0.02&\\
ClassII107&18:03:02.76-22:58:52.0& 3.76$\pm$ 0.04& 8.16$\pm$ 0.14& 7.93$\pm$ 0.03& 8.13$\pm$ 0.01& 8.22$\pm$ 0.01& 9.72$\pm$ 0.02&99.99$\pm$ 9.99&14.60$\pm$ 0.02&\\
ClassII108&18:03:03.38-22:57:19.1& 3.91$\pm$ 0.03& 7.94$\pm$ 0.05& 7.85$\pm$ 0.02& 8.27$\pm$ 0.01& 8.19$\pm$ 0.01& 9.24$\pm$ 0.02&99.99$\pm$ 9.99&13.57$\pm$ 0.02&\\
ClassII109&18:03:03.50-22:57:03.6& 4.23$\pm$ 0.06& 7.76$\pm$ 0.05& 8.25$\pm$ 0.02& 8.45$\pm$ 0.01& 8.76$\pm$ 0.01&10.26$\pm$ 0.02&99.99$\pm$ 9.99&15.96$\pm$ 0.02&\\
ClassII110&18:03:03.82-22:57:34.9& 3.75$\pm$ 0.02& 7.11$\pm$ 0.03& 7.30$\pm$ 0.01& 7.74$\pm$ 0.01& 7.67$\pm$ 0.01& 8.85$\pm$ 0.02&99.99$\pm$ 9.99&13.56$\pm$ 0.02&\\
ClassII111&18:03:04.27-23:08:02.4& 4.97$\pm$ 0.04& 8.04$\pm$ 0.02& 9.41$\pm$ 0.03& 9.76$\pm$ 0.03&10.07$\pm$ 0.03&12.13$\pm$ 0.02&99.99$\pm$ 9.99&14.26$\pm$ 0.02&\\
\enddata
\end{deluxetable}

\begin{deluxetable}{llllccccccccc}
\tabletypesize{\scriptsize}
\rotate
\tablenum{2}
\tablewidth{0pt}
\tablecaption{Properties of Protostars within the Four Dust Continuum Cores}.
\label{proto}
\tablehead{
\colhead{Source}    &
\colhead{   SSTM20  }    &
\colhead{   MIPS 24 }    &
\colhead{   IRAC 8   }    &
\colhead{   IRAC 5.8  }    &
\colhead{  IRAC 4.5 }    &
\colhead{IRAC 3.6  }    &
\colhead{K\tablenotemark{a}}    &
\colhead{H\tablenotemark{a} }    &
\colhead{J\tablenotemark{a}}    &
\colhead{sep.\tablenotemark{b}}    &
\colhead{  L_{mid-IR}}  &
\colhead{  } }
\tablehead{
\colhead{ name   }       & 
\colhead{   (J2000)  }              &
\colhead{    (mag) } &
\colhead{    (mag)} & 
\colhead{    (mag)} & 
\colhead{    (mag)} & 
\colhead{    (mag)} &     
\colhead{(mag)} & 
\colhead{(mag)}& 
\colhead{(mag)} & 
\colhead{($''$)} &
\colhead{ (L$_\odot$)}  &
\colhead{ }  
 }     
\startdata
TC3A&18:02:05.57-23:05:29.0& 2.93 0.03& 7.88 0.08& 8.45 0.03& 9.66 0.02&11.18 0.04&99.99 9.99&18.04 0.40&99.99 9.99& 0  & 13& \\
TC3B&18:02:07.20-23:05:36.2& 3.94 0.08& 7.18 0.07& 7.54 0.03& 8.30 0.01& 9.44 0.02&14.43 0.02&99.99 9.99&99.99 9.99& 23.6&9 &  \\
TC3C&18:02:05.45-23:05:05.3& 4.36 0.10& 7.92 0.12& 8.71 0.05& 9.57 0.02&11.34 0.05&16.49 0.08&99.99 9.99&99.99 9.99& 23.8& 4 & \\
TC3D&18:02:05.52-23:04:39.4& 7.34 0.59& 9.59 0.01& 9.49 0.05&10.10 0.03&11.57 0.05&99.99 9.99&20.07 0.78&20.11 0.42& 49.7& 1 &  \\
TC3E&18:02:01.77-23:05:53.9& 3.03 0.02& 7.13 0.02& 7.71 0.01&99.99 9.99& 9.54 0.02&12.91 0.02&15.73 0.05&19.09 0.18& 57.9 &16 &  \\
TC4A&18:02:12.77-23:05:46.7& 1.83 0.02& 5.92 0.01& 6.77 0.01& 8.25 0.01& 9.87 0.02&14.23 0.02&15.61 0.06&17.17 0.04& 0    &32 & \\
TC4B&18:02:13.08-23:06:07.2& 3.89 0.11& 8.59 0.10& 9.19 0.04& 9.75 0.03&11.12 0.04&15.46 0.05&99.99 9.99&99.99 9.99& 21.0& 5&  \\
TC4C&18:02:12.50-23:05:15.4& 4.88 0.28& 6.75 0.01& 7.47 0.01& 8.37 0.01& 9.18 0.02&12.03 0.02&13.91 0.02&16.28 0.02& 31.5& 9  & \\
TC4D&18:02:14.04-23:06:40.7& 3.47 0.09& 8.16 0.11& 8.74 0.06&10.15 0.04&11.69 0.08&16.71 0.15&99.99 9.99&19.53 0.31& 56.8& 6.2& \\
TC4E&18:02:15.79-23:06:42.8& 3.57 0.10& 7.51 0.06& 7.47 0.01& 8.26 0.01& 9.45 0.02&15.06 0.02&18.12 0.44&99.99 9.99& 70.0& 10 & \\
TC1 &18:02:24.68-23:01:17.7& 2.02 0.90& 8.87 0.55& 9.19 0.17& 9.70 0.06&10.68 0.11&17.24 0.35&17.91 0.29&19.43 0.27& --& 19 \\
TC2 &18:02:28.49-23:03:56.9& 2.09 0.11& 5.76 0.02& 7.69 0.03&10.48 0.06&10.78 0.07&13.57 0.02&16.20 0.08&19.15 0.20& --& 24\\
\enddata
\tablenotetext{a}{near-infrared data are from Palomar 200 inch measurements.}
\tablenotetext{b}{separation of TC3B-TC3E  from the central protostar TC3A in the TC3 core,
and separation of TC4B-TC4E from the central protostar TC4A in the TC 4 core.
for TC4 condensation.}
\tablenotetext{c,d}{L$_{mid-IR}$ is mid-infrared luminosity estimated from 1.2-24$\mu$m.}
\end{deluxetable}

\normalsize

{}

\clearpage
\begin{figure}
\caption{
Mosaicked three-color {\it Spitzer} image of the Trifid Nebula. Blue,
green and red represent IRAC 4.5 and 8$\mu$m, and MIPS 24$\mu$m images,
respectively. PAH-dominated emission appears in green, hot dust grains
appear in red, and young stellar objects appear as point sources in red or yellow.
Also note the filamentary dark clouds  on the western side of M20. The diffuse
emission at 8 and 24$\mu$m ranges between 80 and 450 MJy sr$^{-1}$, and
between 90 and 280 MJy sr$^{-1}$, respectively. The image is centered at R.A.\
$18^{\rm h} 02^{\rm m} 26^{\rm s}$ and Dec.\ $-23^\circ$00$^{\prime}
39^{\prime \prime}$ (J2000), and covers  an 18$'$$\times$26$'$ arcmin field
of view. {\it (Fig. 1 is a jpeg file) } }
\label{trispicolor}
\end{figure}

\begin{figure}
\psfig{figure=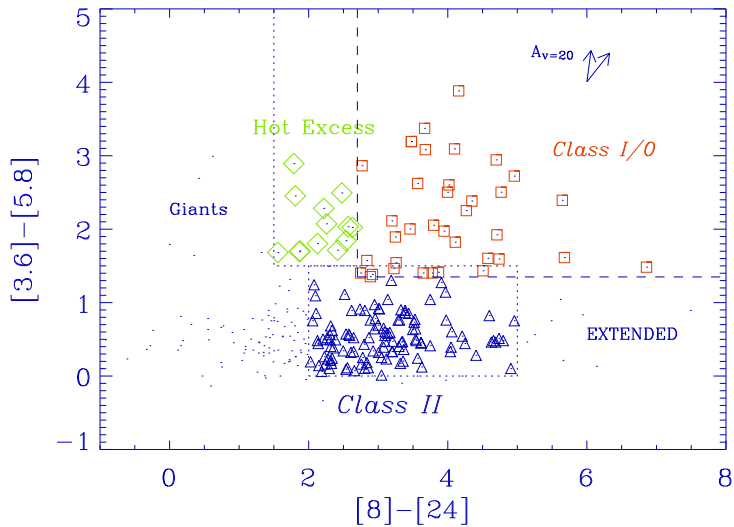,width=11truecm,angle=0}
\caption{\spitzer\ color-color diagram in the Trifid Nebula, using  IRAC and MIPS data.
The  approximate locations of Classes of I/0,  II and ``Hot excess" 
YSOs are outlined. Protostars of Classes I/0 are the reddest stars (squares) located in the
upper right in this diagram, and evolved YSOs of Class II
are red in the color of [8]-[24] and marked with triangles.
The extinction vectors for R$_v$ = 5.5 (thin line) and  R$_v$ = 3.1 (thick line) are shown 
are arrows, which are estimated based on Weingartner \& Draine (2001).
The extinction vector by using other set of \spitzer\ IRAC data \citep{ind05}
is approximately consistent with the case of R$_v$ = 5.5.  
}
\label{tricolcol}
\end{figure}

\begin{figure}
\psfig{figure=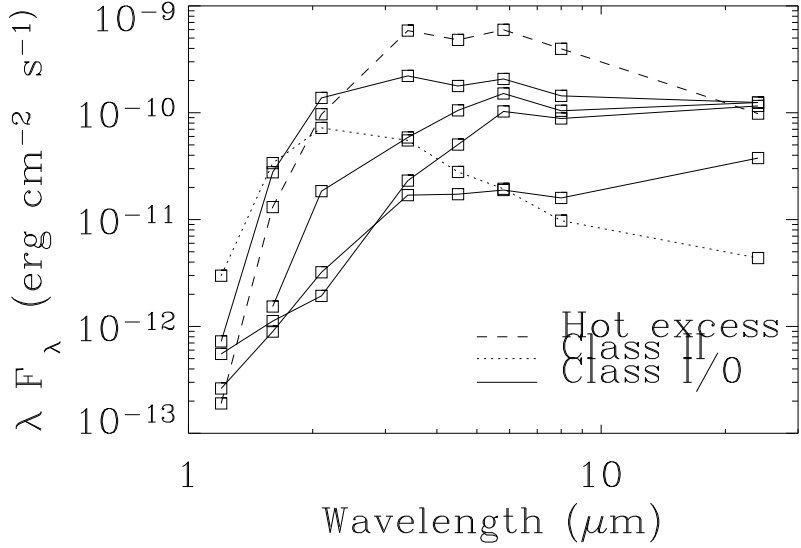,width=12truecm,angle=0}
\caption{ Examples of near- and mid-infrared spectral energy distributions of three different
Classes of YSOs:
Class I/0, Class II and ``Hot excess" stars.
}
\label{classsed}
\end{figure}

\begin{figure}
\psfig{figure=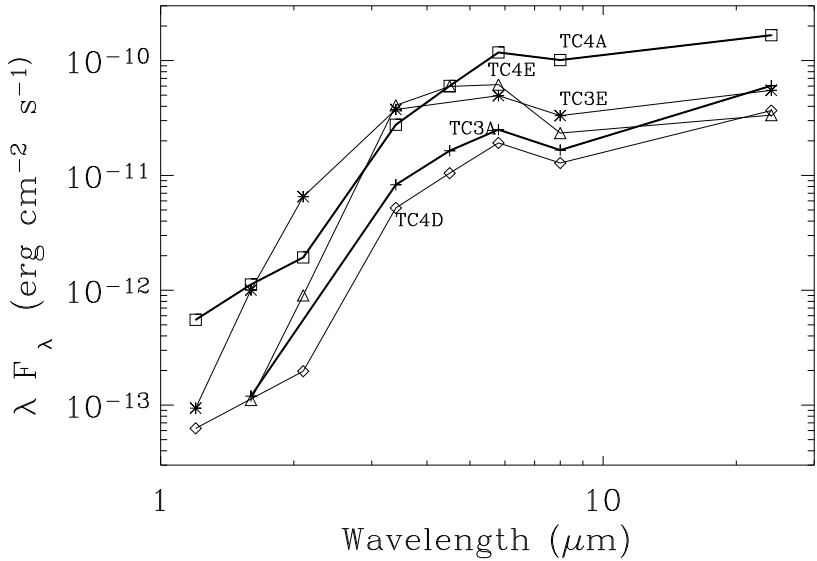,width=12truecm,angle=0}
\caption{Near- and Mid-infrared spectral energy distributions of TC3A (crosses), TC3E (astericks),
TC4A (squares), TC4D (diamonds), and TC4E (triangles). The sources show dips at 8$\mu$m, which is
likely due to  silicate absorption feature.}
\label{sedtc43}
\label{classsedmidir}
\end{figure}

\begin{figure}
\caption{Comparison of the HST (on left) and {\it Spitzer} IRAC
(on right)  images of the Trifid Nebula. While the protostars
appear dark in the optical HST image, they  appear as infrared color
excess stars (yellow or green) in the {\it Spitzer} images. Four
condensations are marked as circles for TC1 (top) and 
TC2 (middle), and ellipes  for TC3 (bottom right) and TC4
(bottom left), and the protostars within the cores are marked as 
arrows.  Four of the five protostars within each of TC3 and TC4 cores are
marked, because they are clearly noticeable in the IRAC color images; 
all five protostars are shown in Figure \ref{tc3tc4spitzer} (see the
text for details). North is up and west is to the right. {\it (Fig. 5 is a jpeg file)}}
\label{hstspitzer}
\end{figure}

\begin{figure}
\caption{A portion of the {\it Spitzer} color image (from Fig. 1) 
superposed on 1300$\mu$m contours from \cite{lef00}.
The contour levels are 5, 10, 15, 20, 30, 45, 60 by 30, 
200 to 350 Jy by 50 mJy beam$^{-1}$, where 
the beam size is 11$''$. 
Class I/0 sources are marked as green circles, and Class II and $``$hot
excess" sources are marked as red and white circles. The TC3 and TC4
mm sources are each associated with multiple protostars. The brightest
protostars appear near the centers of the dust continuum peaks. 
 {\it (Fig. 6 is a jpeg file)}}
\label{tc3tc4spitzer}
\end{figure}

\begin{figure}
\psfig{figure=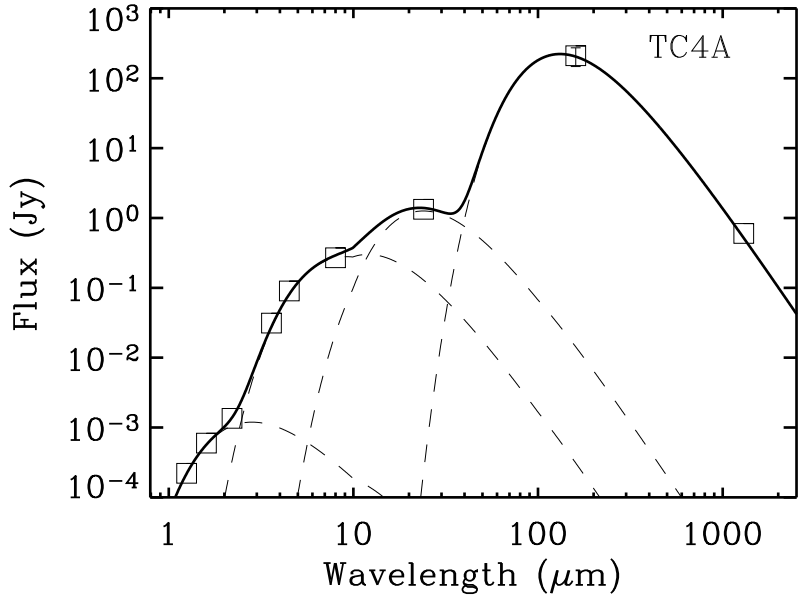,width=12truecm,angle=0}
\caption{ Broad band spectral energy distribution of TC4A, the most massive protostars of TC4.  
The best-fit black body model with four temperatures of T$_{cold}$=22 K,
T$_{warm}$=150 K, T$_{hot1}$=400 K, and T$_{hot2}$= 1300 K  is shown.} 
\label{sedtc4a}

\end{figure}



\end{document}